\documentclass[]{emulateapj-rtx4}

\usepackage{natbib,color} 
\bibpunct{(}{)}{;}{a}{}{,}

\DeclareRobustCommand{\ion}[2]{%
\relax\ifmmode
\ifx\testbx\f@series
{\mathbf{#1\,\mathsc{#2}}}\else
{\mathrm{#1\,\mathsc{#2}}}\fi
\else\textup{#1\,{\mdseries\textsc{#2}}}%
\fi}

\shorttitle{Thermodynamic Properties of the Inverse Evershed Flow}
\shortauthors{Choudhary, D.P.; Beck, C.}

\begin{document}
\title{Thermodynamic Properties of the Inverse Evershed Flow in the Lower Chromosphere}

\author{D.P. Choudhary}
\affil{Department of Physics \& Astronomy, California State University, Northridge, CA}
\author{C. Beck}
\affil{National Solar Observatory, Boulder, CO}




\begin{abstract}
We used spectropolarimetric observations of a sunspot in active region NOAA 11809 in the \ion{Ca}{ii} line at 854.2\,nm taken with the SpectroPolarimeter for Optical and Infared Regions (SPINOR) at the Dunn Solar Telescope to infer thermodynamic parameters along one hundred super-penumbral fibrils that harbor the inverse Evershed flow. The fibrils were identified in line-of-sight (LOS) velocity and line-core intensity maps and were located in a segment of the sunspot that showed a regular penumbra in the photosphere. The chromospheric LOS velocity abruptly decreases from 3--15 km\,s$^{-1}$ to zero at the inner footpoints of the fibrils that are located from the mid penumbra to about 1.4 spot radii. The spectra often show multiple components, i.e., one at the rest wavelength and one with a strong red shift, which indicates spatially or vertically unresolved structures. The line-core intensity always peaks slightly closer to the umbra than the LOS velocity. An inversion of the spectra with the CAlcium Inversion using a Spectral ARchive (CAISAR) code provided us with temperature stratifications that allowed us to trace individual fibrils through the atmosphere and to determine the angle of the flows relative to the surface without any additional assumptions on the flow topology such as radial symmetry. We find that the fibrils are not horizontal near the downflow points, but make an angle of 30--60 degrees to the local vertical. The temperature is enhanced by 200\,K at log\,$\tau \sim -2$ and up to 2000\,K at log\,$\tau \sim -6$ over that of the quiet Sun, whereas there is no signature in the low photosphere. Our results are consistent with a critical, i.e., sonic, or super-sonic siphon flow along super-penumbral flux tubes in which accelerating plasma abruptly attains sub-critical velocity through a standing shock in or near the penumbra as predicted by \citet{montesinos+thomas1993aa}. 
\end{abstract}

\section{Introduction}
The mass motions inside and in the vicinity of sunspots are mostly organized as flows along penumbral filaments in the photosphere and along super-penumbral fibrils in the chromosphere. The underlying magnetic field topology is assumed to consist of a complex mixture of vertical, partially inclined or horizontal field lines \citep[e.g.,][]{solanki2003,beck2008,rempel+schlichenmaier2011}. The direction and strength of the flows along different channels depends on the thermodynamic and magnetic properties of the regions connected by the magnetic field lines. The determination of the flow speed, the flow direction and the temperature of the plasma in these flow channels is critical for understanding the underlying physical mechanisms that are to a large extent governed by the global magnetic structure. 

Early spectroscopic observations of sunspots revealed a peculiar flow pattern with a radial, horizontal outflow at photospheric layers, the Evershed effect \citep{evershed1909}. At high spatial resolution, the Evershed effect was found to be structured into thin, elongated radial flow filaments in which the material flows intermittently at a speed of about 1--6 km\,s$^{-1}$ along those elevated flow channels from inside the sunspot to the end of the penumbra and beyond \citep[]{ichimoto+etal2007,bumba1960,reza+etal2006,rimmele+marino2006}. The flow speed decreases at large radial distances from the sunspot center due to the development of a shock front \citep{borrero+etal2005}. At chromospheric heights, mass motions with higher speeds were observed along the so-called super-penumbral fibrils. The chromospheric mass propagates in a direction opposite to the photospheric flow \citep[``inverse'' Evershed flow;][]{evershed1910,stjohn1911,stjohn1911a,maltby1975,tsiropoula2000}. In H$\alpha$ dopplergrams, flow speeds in the range of 20--50 km\,s$^{-1}$ were found \citep{beckers1964,moore1981}. On the other hand, spectroscopic measurements showed lower speeds in the range of 3-–7 km\,s$^{-1}$ \citep{haugen1969,dialetis+etal1985,dere+etal1990}. Dark super-penumbral fibrils exhibited a faster flow velocity than bright ones, with fluctations on a time scale of about 25 min \citep{georgakilas+etal2003,georgakilas+christopoulou2003}. 

The inverse Evershed flow was found to start with subsonic speeds away from the sunspot, to attain super-sonic speeds near the loop top and to end with subsonic speeds inside the sunspot by \citet{maltby1975}. Wavelength shifts in the core and wing of the \ion{Mg}{i} b1 518.362 nm line showed a reverse flow at photospheric and chromospheric heights \citep{bones+maltby1978}. Simultaneous measurements of Doppler velocities usually show an increase in the flow speed from the photosphere through the chromosphere to transition region heights \citep{alissandrakis+etal1988,bethge+etal2012}. The steady flow pattern in Dopplergrams of the \ion{C}{iv} transition-region line at 154.8\,nm observed with the Solar Maximum Mission was seen to be consistent with the inverse Evershed effect \citep{athay+etal1982}.

The photospheric Evershed effect could be due to a siphon flow mechanism \citep{montesinos+thomas1997,sanchezalmeida+etal2007} or could be a manifestation of magnetoconvection in the penumbra \citep{rempel2012}, which is a question of considerable debate in the recent years. For the chromospheric inverse Evershed effect it is, however, generally accepted that it is caused by siphon flows along arched magnetic field lines connecting footpoints with a different magnetic field strength and gas pressure \citep[][]{meyer+schmidt1968,thomas1988,degenhardt+etal1993,thomas+montesinos1993}. \citet{thomas1988} and the subsequent papers of the series \citep{montesinos+thomas1989,thomas+montesinos1990,thomas+montesinos1991,montesinos+thomas1993aa} developed an extensive description of such siphon flows in the solar photosphere. One major distinction they found was related to the question whether the flow speed was below (sub-critical), at (critical) or above (super-critical) the local sound speed. They predicted the generation of a shock front wherever the flow speed reached the super-critical level. Observational evidence for the existence of such shock fronts was found both at photospheric \citep{degenhardt+etal1993} and chromospheric heights \citep{uitenbroek+etal2006,beck+etal2010a,bethge+etal2012}. 

In this paper, we derive the thermodynamic properties of the solar atmosphere from an analysis of \ion{Ca}{ii} IR spectra at 854.2\,nm to determine the properties of the inverse Evershed flow in the lower chromosphere. Section \ref{secobs} describes our set of observations. Section \ref{secanalysis} explains the retrieval of atmospheric properties from the data. The results are presented, summarized and discussed in Sections \ref{secres} to \ref{secdisc}. Section \ref{secconcl} provides our conclusions.
\section{Observations}\label{secobs}
We observed the leading sunspot of the active region (AR) NOAA 11809 on 3 August 2013 with the \textit{SPectropolarimeter for Infrared and Optical Regions} \citep[SPINOR;][]{socasnavarro+etal2006} at the Dunn Solar Telescope \citep[DST;][]{dunn1969,dunn+smartt1991}. The sunspot was located at an heliocentric angle of about 30 degree. We acquired spectropolarimetric data in the chromospheric \ion{Ca}{ii} infrared (IR) line at 854.2\,nm and the photospheric \ion{Fe}{i} lines at 1564.8\,nm and 1565.2\,nm. Between UT 15:24 and UT 18:43, in total nine scans of the sunspot were taken with a cadence of about 30 min. For the first map, 400 steps with a step width of 0\farcs22 were recorded while all other scans had 200 steps. The integration time per step was 4\,s. The slit width was 30\,$\mu$m corresponding to about 0\farcs22 on the Sun. The spatial (spectral) sampling along the slit was 0\farcs36 (5.5\,pm) at 854\,nm and 0\farcs55 (20.6\,pm) at 1565\,nm. The field of view (FOV) along the slit was about 90$^{\prime\prime}$ at 854\,nm  and about 150$^{\prime\prime}$ at 1565\,nm. The full wavelength range covered was 852.87--855.70\,nm and 1558.05--1568.58\,nm, respectively. The spectropolarimetric data were reduced with the standard SPINOR data reduction pipeline\footnote{\url{http://nsosp.nso.edu/dst-pipelines}}. 

\begin{figure}
\centerline{\resizebox{6cm}{!}{\includegraphics{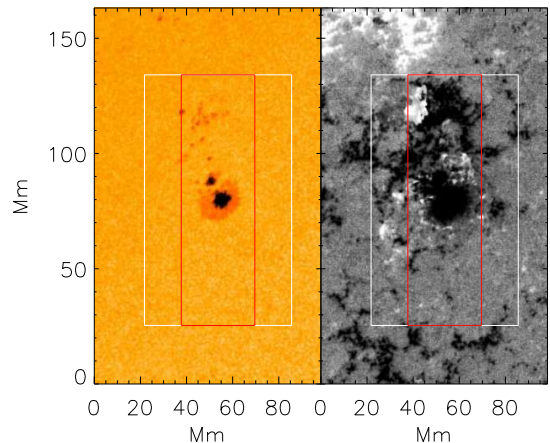}}}$ $\\$ $\\
\caption{Overview of the FOV in NOAA 11809 on 2013/08/03 in HMI data. Left: continuum intensity. Right: LOS magnetic flux. The white and red rectangles indicate the areas scanned in the first and all other maps, respectively.}\label{fig1}
\end{figure}

Figure \ref{fig1} shows an overview of the observed FOV in the continuum intensity and the magnetogram from the \textit{Helioseismic and Magnetic Imager} \citep[HMI;][]{scherrer+etal2012} onboard  the \textit{Solar Dynamics Observatory} \citep[][]{pesnell+etal2012}. The leading sunspot of NOAA 11809 had negative polarity and belonged to a decaying active region. The trailing polarity was located near the upper left corner of the FOV shown, while the several small pores next to the sunspot at $x,y \sim 40-60$\,Mm, $100-130$\,Mm belonged to newly emerging magnetic flux. 
\section{Data Analysis}\label{secanalysis}
We derived an extended list of line parameters from the intensity and polarization profiles in both wavelength ranges. In the current study, we only use the continuum intensity $I_c$, the line-core intensity $I_{\rm core}$ and line-core velocity $v_{\rm core}$ of \ion{Ca}{ii} IR, and the line-core velocity $v_{\rm phot}$ of one photospheric IR line in the 1565\,nm range.

\begin{figure*}
\centering
\resizebox{3cm}{!}{\includegraphics{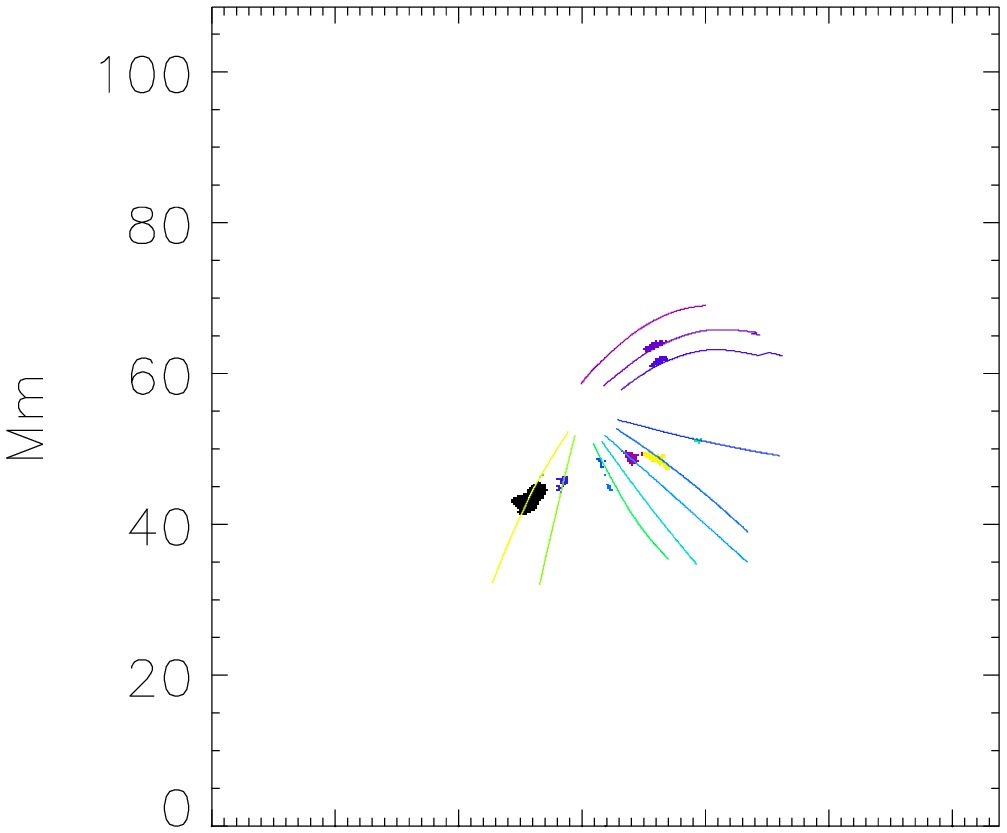}}
\resizebox{1.5cm}{!}{\includegraphics{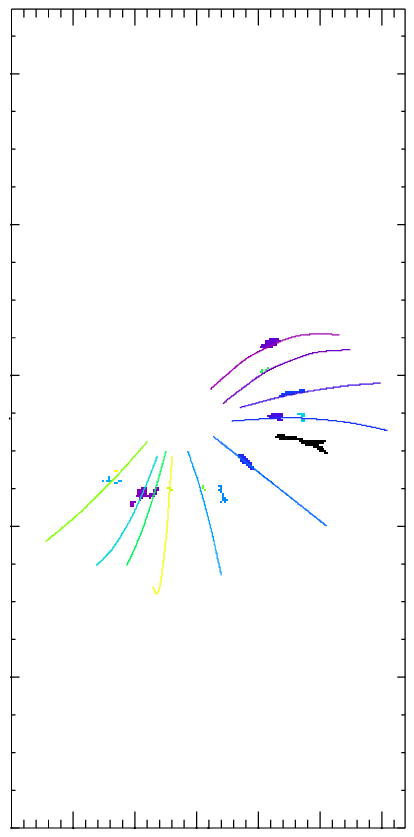}}
\resizebox{1.5cm}{!}{\includegraphics{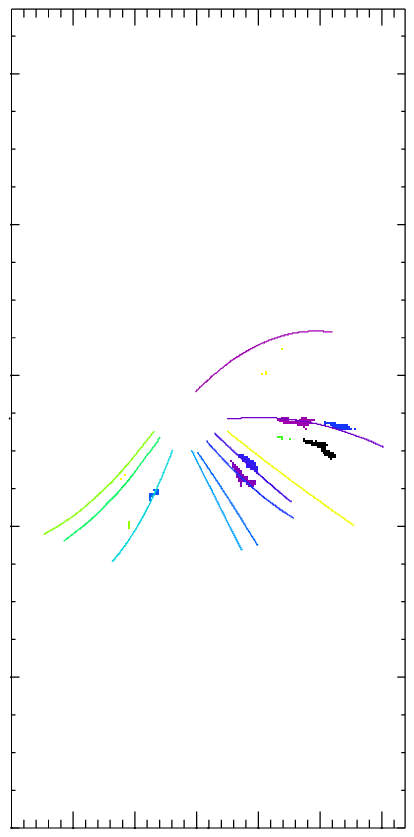}}
\resizebox{1.5cm}{!}{\includegraphics{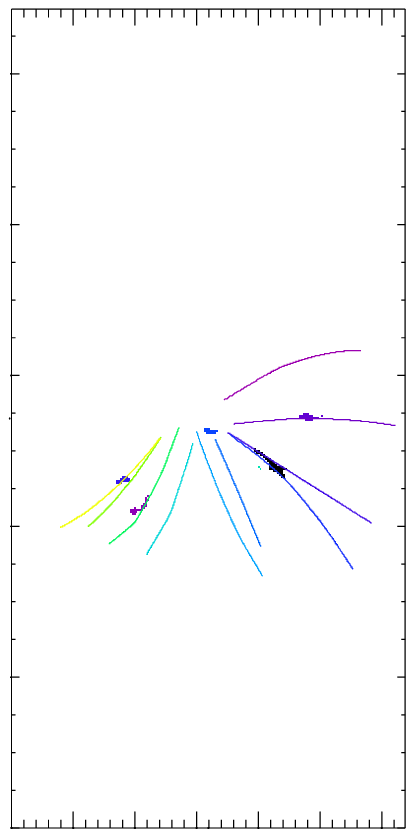}}
\resizebox{1.5cm}{!}{\includegraphics{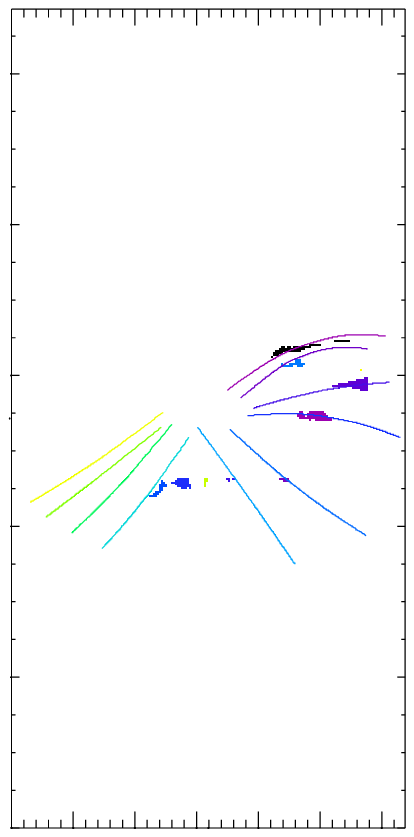}}
\resizebox{1.5cm}{!}{\includegraphics{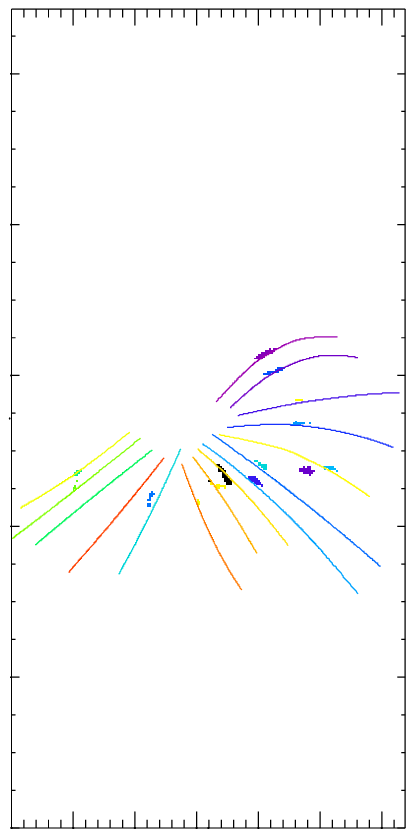}}
\resizebox{1.5cm}{!}{\includegraphics{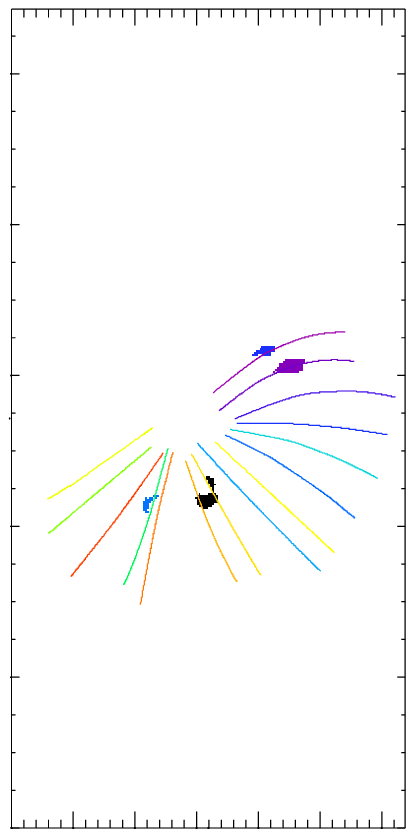}}
\resizebox{1.5cm}{!}{\includegraphics{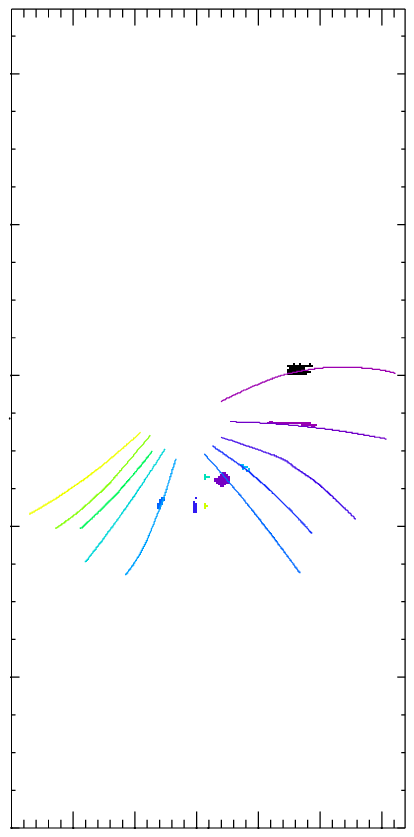}}
\resizebox{1.5cm}{!}{\includegraphics{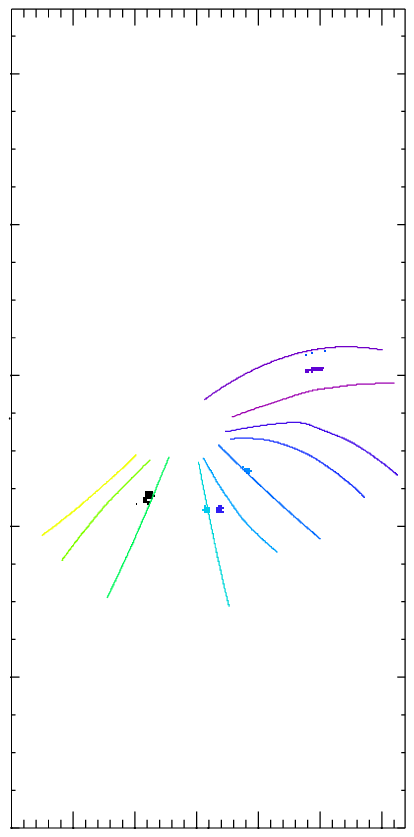}}\\
\resizebox{3cm}{!}{\includegraphics{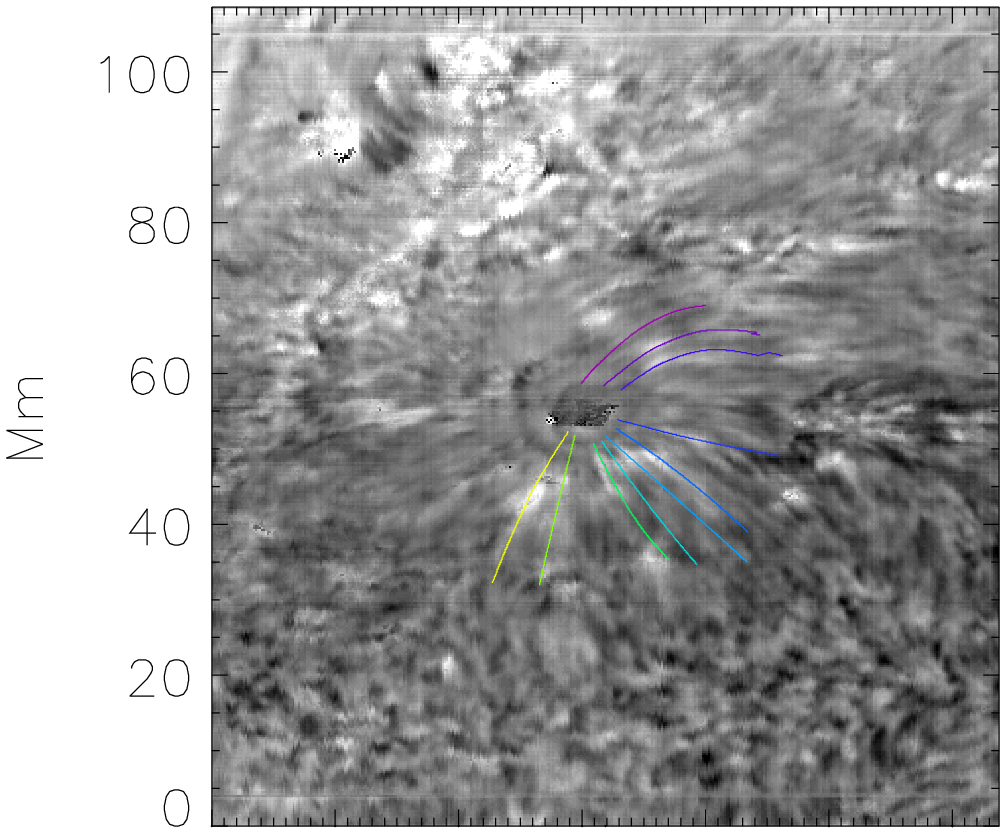}}
\resizebox{1.5cm}{!}{\includegraphics{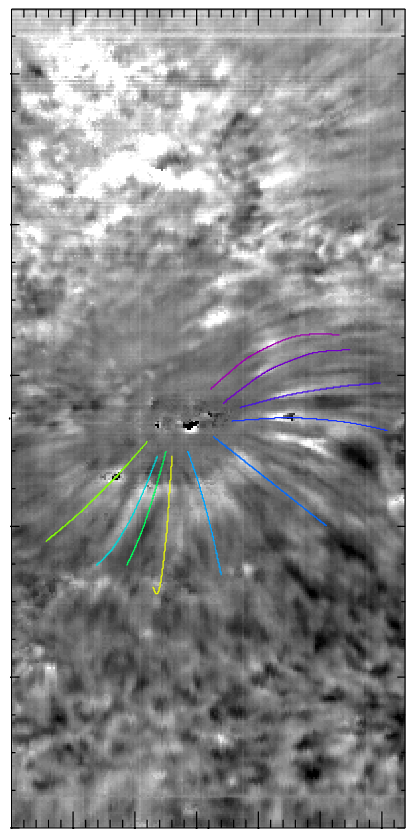}}
\resizebox{1.5cm}{!}{\includegraphics{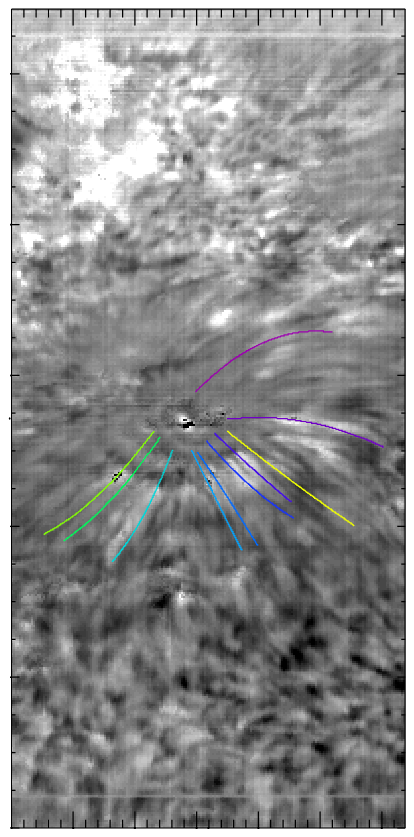}}
\resizebox{1.5cm}{!}{\includegraphics{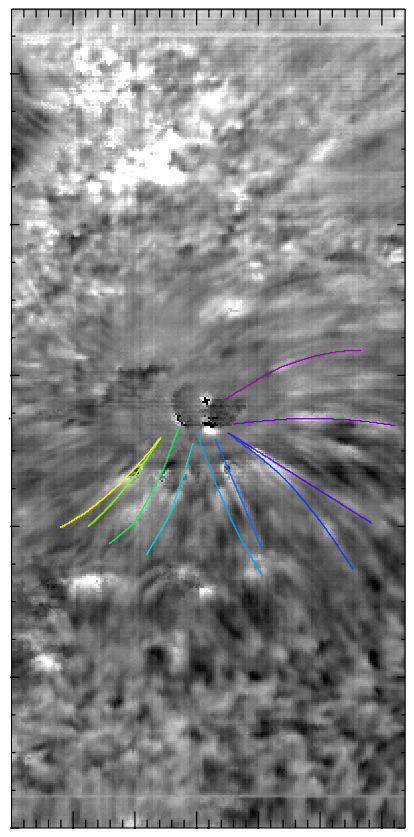}}
\resizebox{1.5cm}{!}{\includegraphics{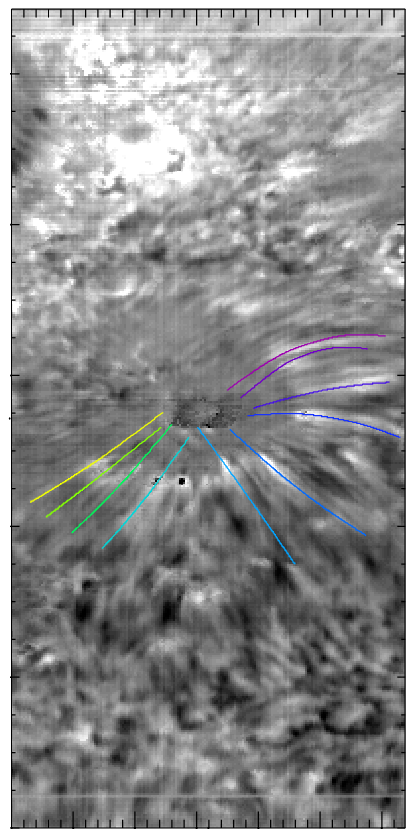}}
\resizebox{1.5cm}{!}{\includegraphics{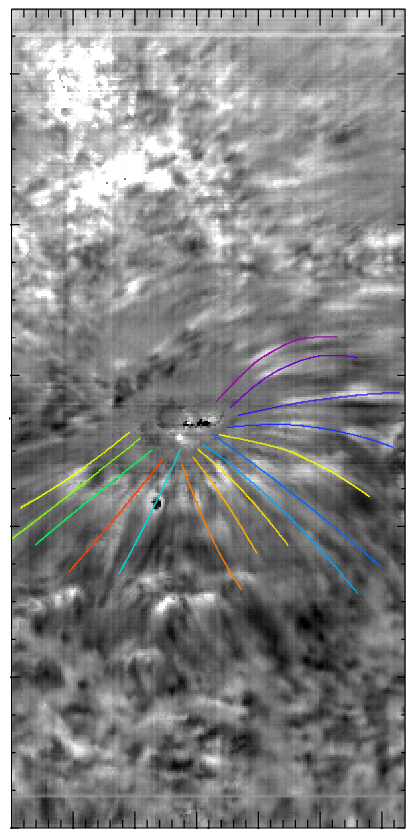}}
\resizebox{1.5cm}{!}{\includegraphics{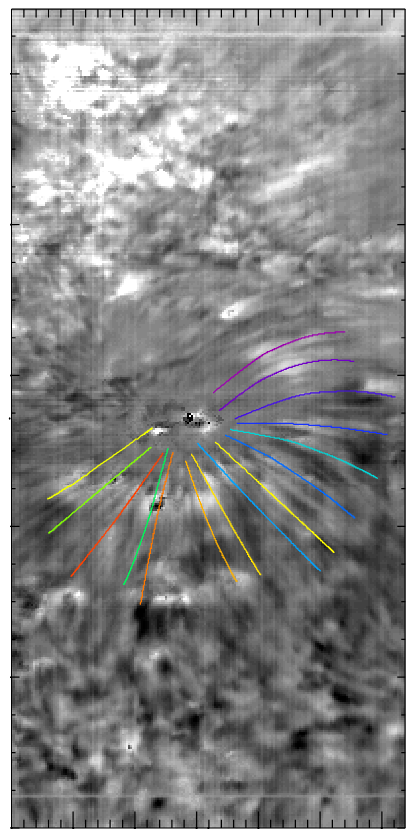}}
\resizebox{1.5cm}{!}{\includegraphics{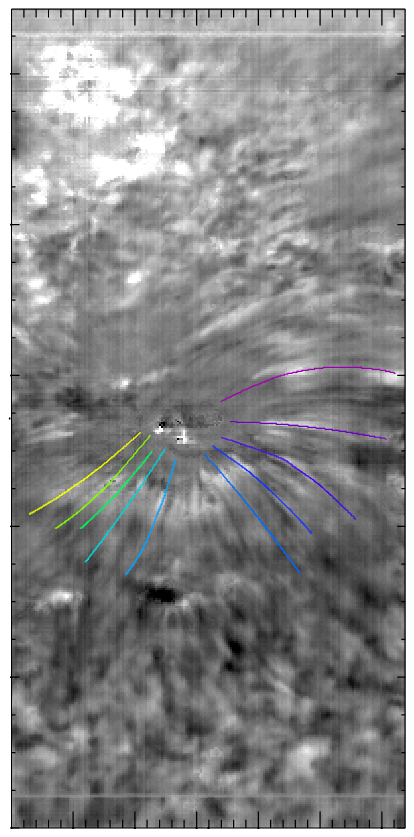}}
\resizebox{1.5cm}{!}{\includegraphics{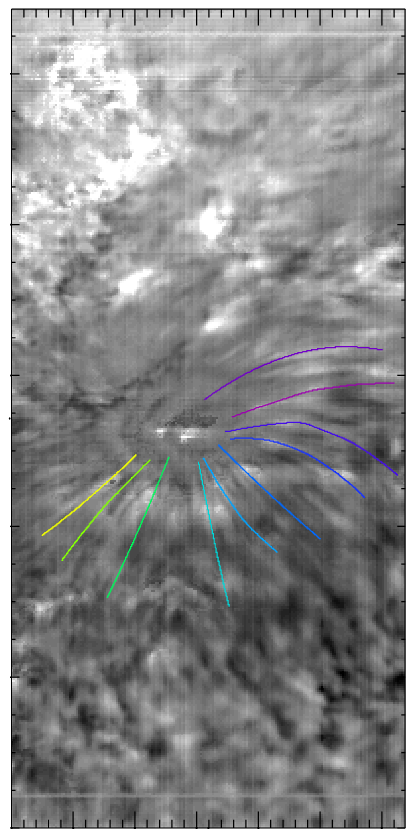}}\\
\resizebox{3cm}{!}{\includegraphics{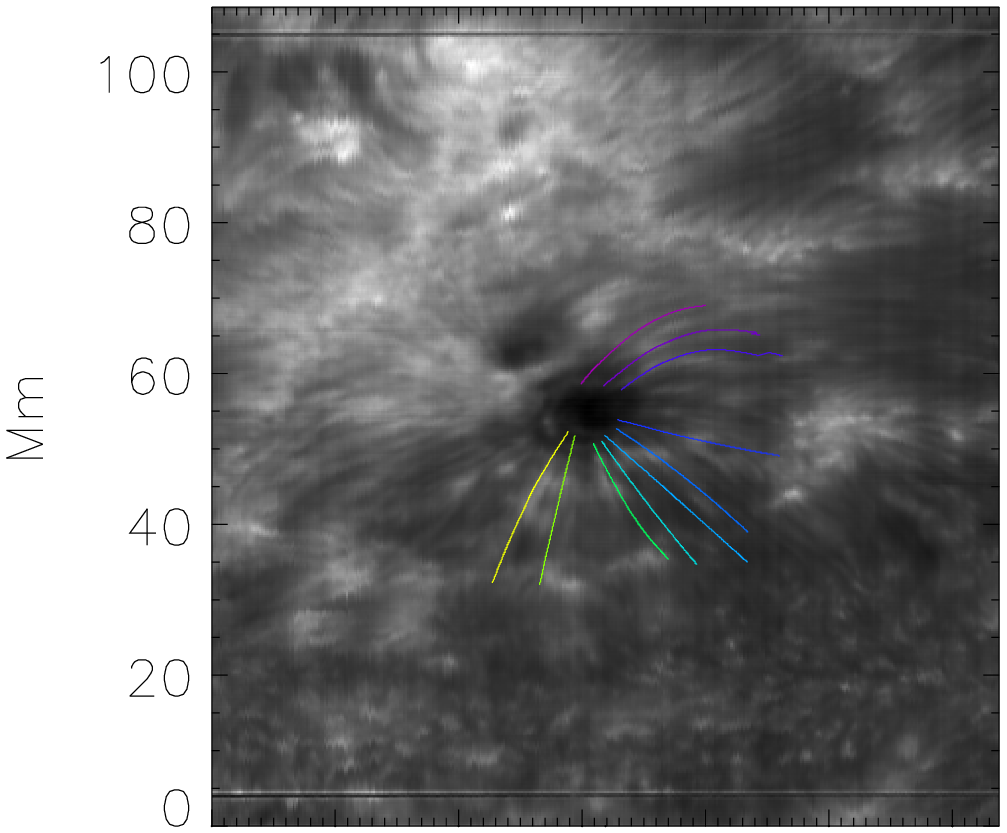}}
\resizebox{1.5cm}{!}{\includegraphics{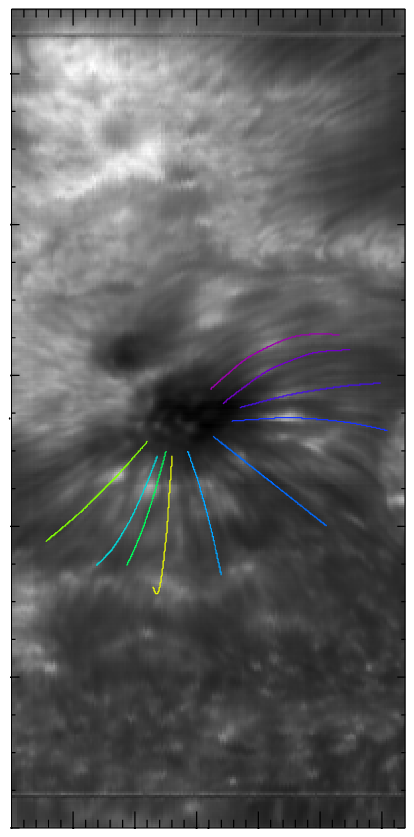}}
\resizebox{1.5cm}{!}{\includegraphics{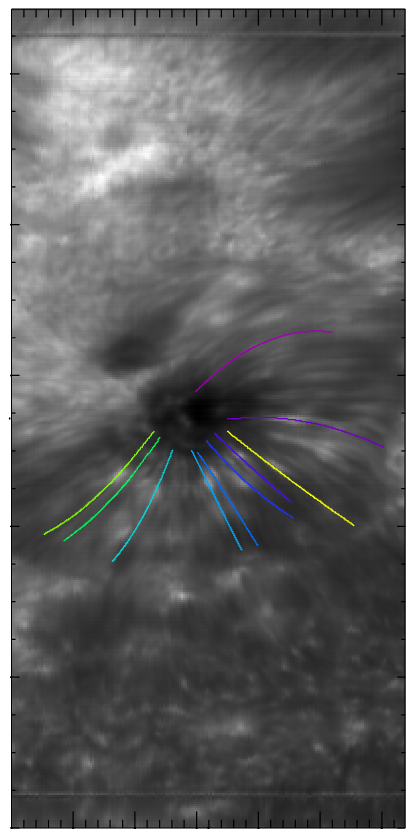}}
\resizebox{1.5cm}{!}{\includegraphics{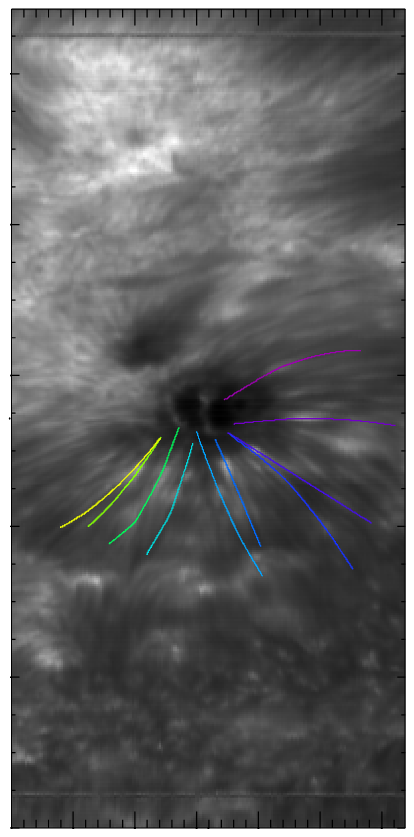}}
\resizebox{1.5cm}{!}{\includegraphics{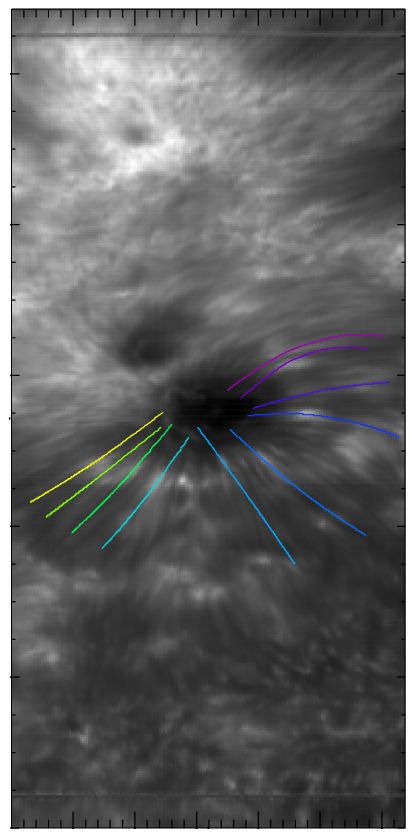}}
\resizebox{1.5cm}{!}{\includegraphics{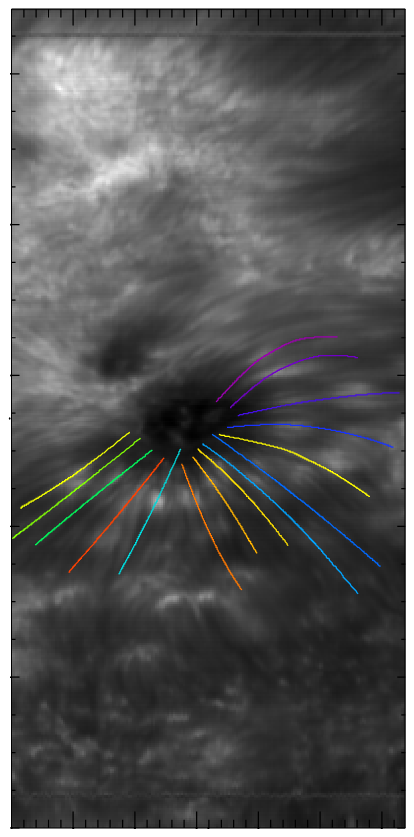}}
\resizebox{1.5cm}{!}{\includegraphics{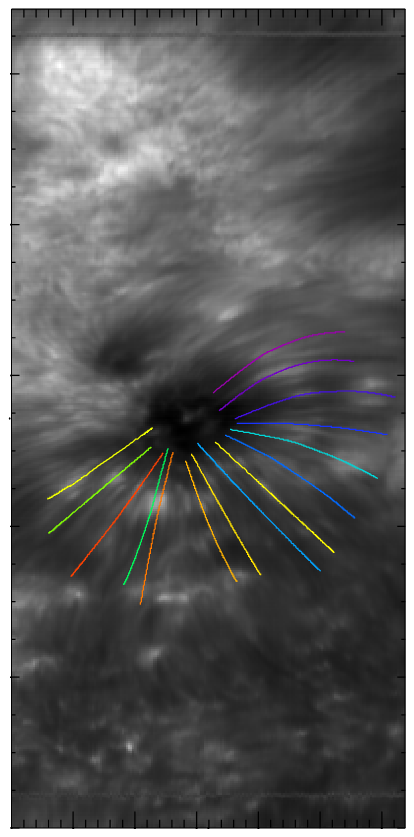}}
\resizebox{1.5cm}{!}{\includegraphics{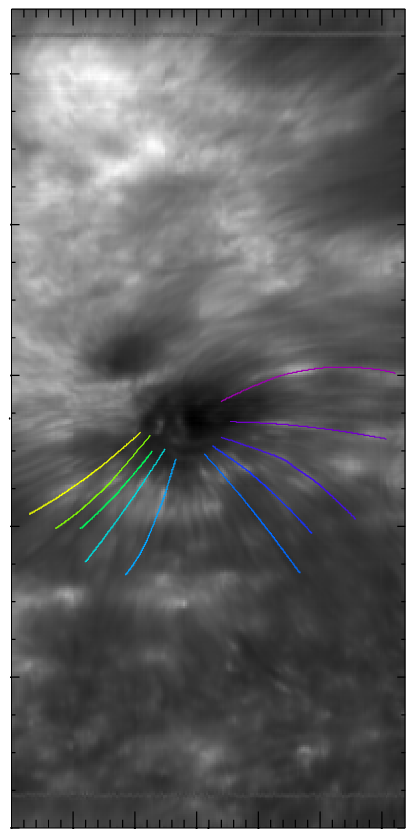}}
\resizebox{1.5cm}{!}{\includegraphics{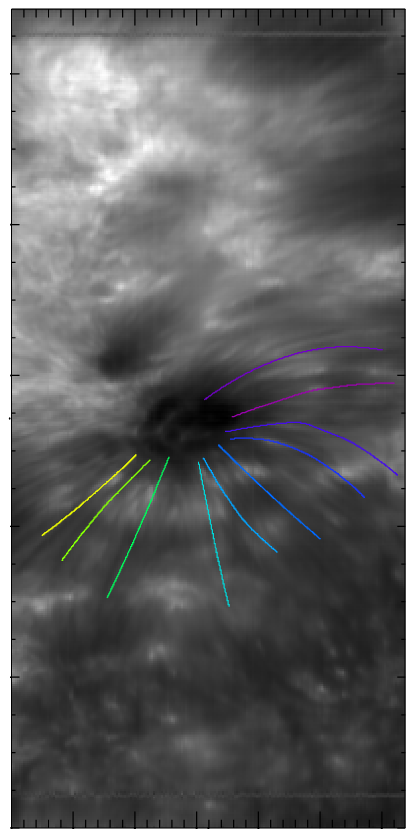}}\\
\resizebox{3cm}{!}{\includegraphics{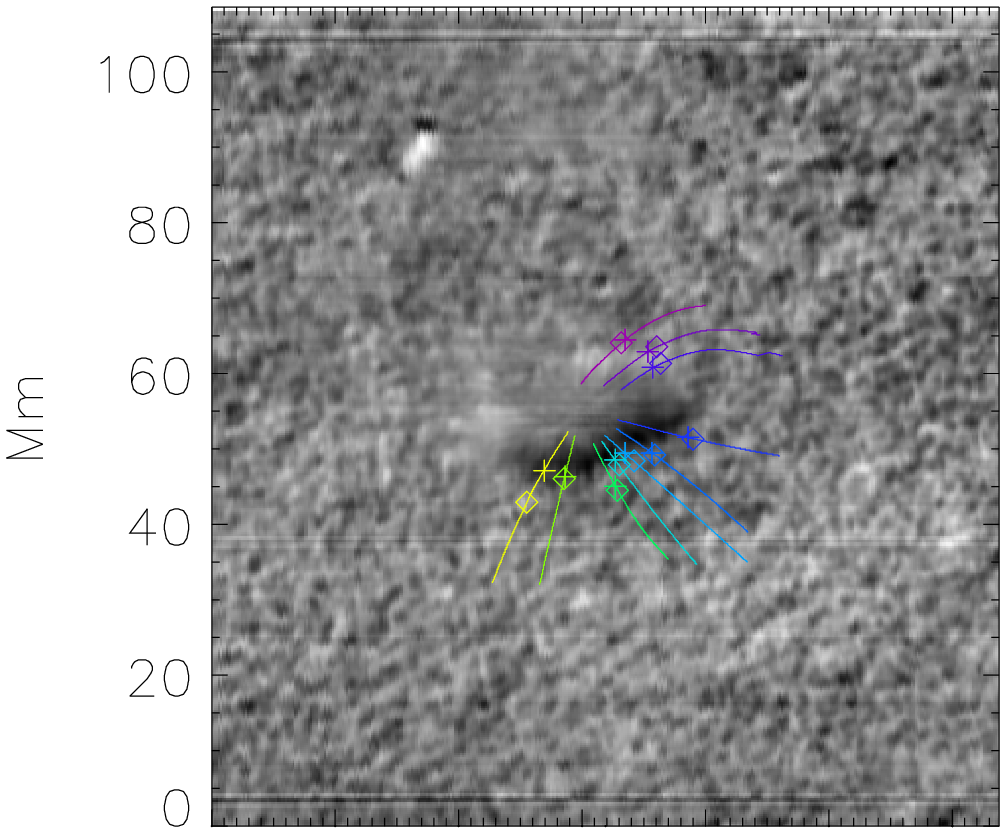}}
\resizebox{1.5cm}{!}{\includegraphics{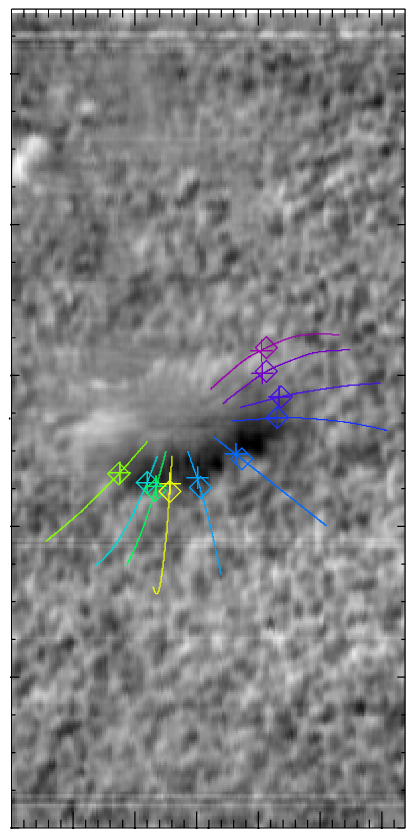}}
\resizebox{1.5cm}{!}{\includegraphics{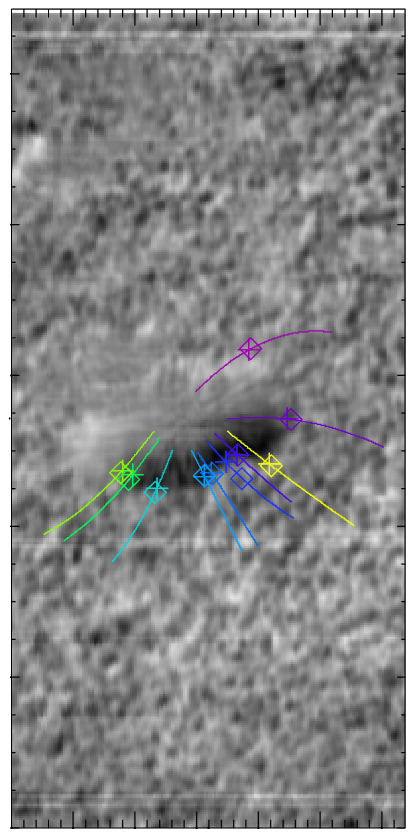}}
\resizebox{1.5cm}{!}{\includegraphics{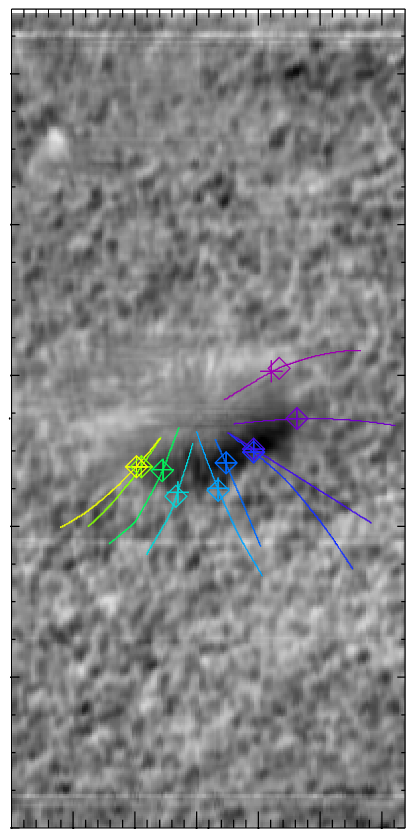}}
\resizebox{1.5cm}{!}{\includegraphics{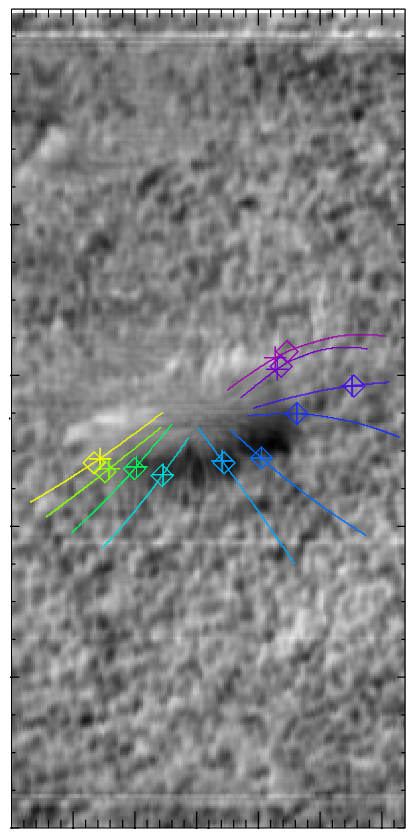}}
\resizebox{1.5cm}{!}{\includegraphics{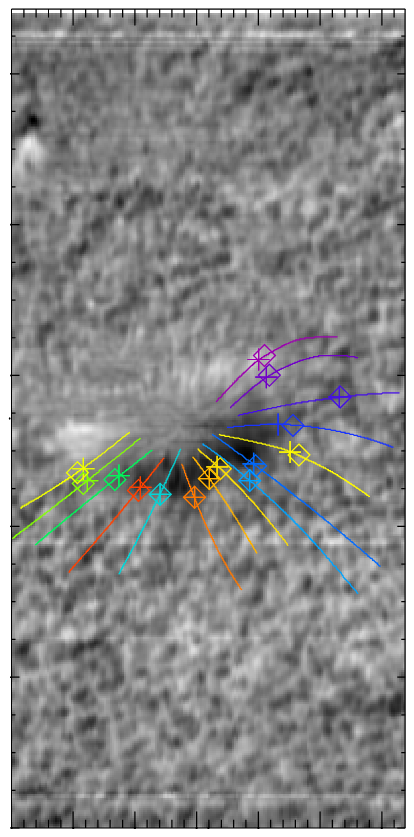}}
\resizebox{1.5cm}{!}{\includegraphics{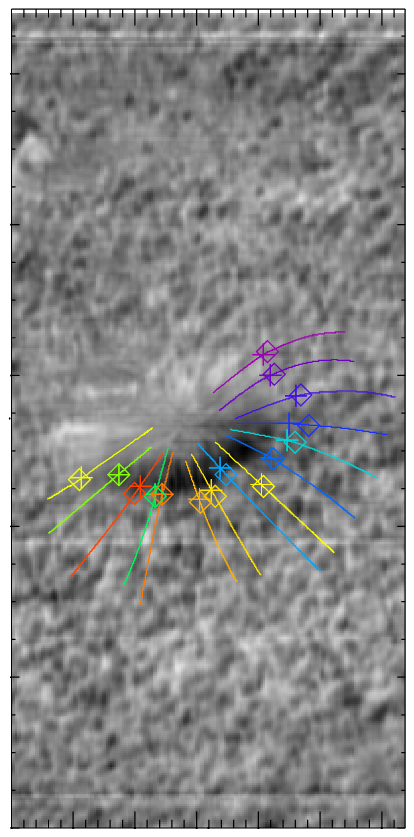}}
\resizebox{1.5cm}{!}{\includegraphics{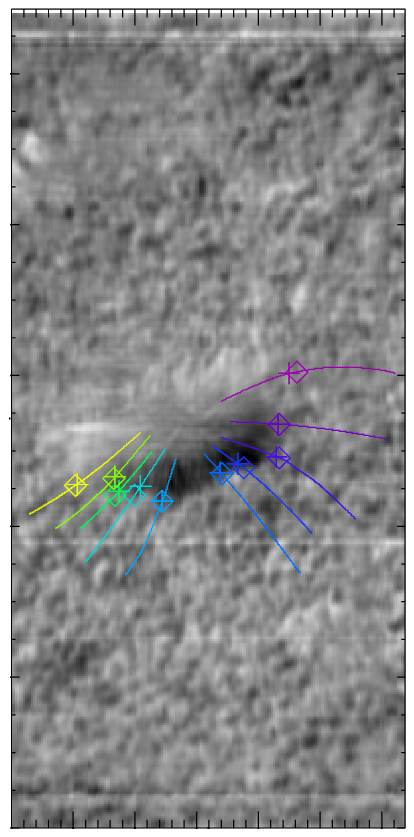}}
\resizebox{1.5cm}{!}{\includegraphics{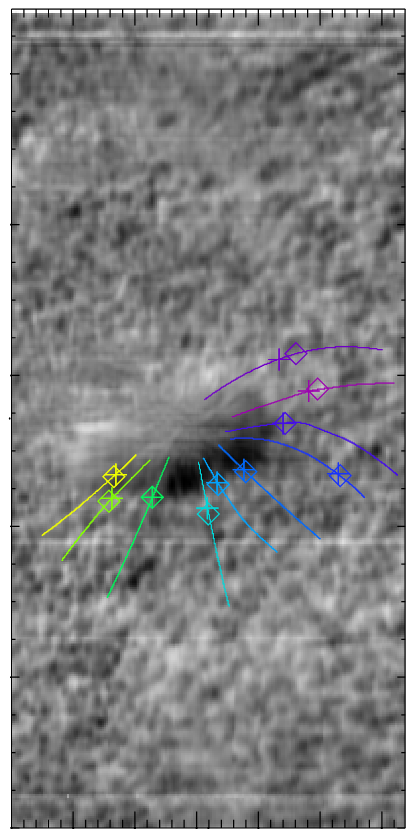}}\\
\resizebox{3cm}{!}{\includegraphics{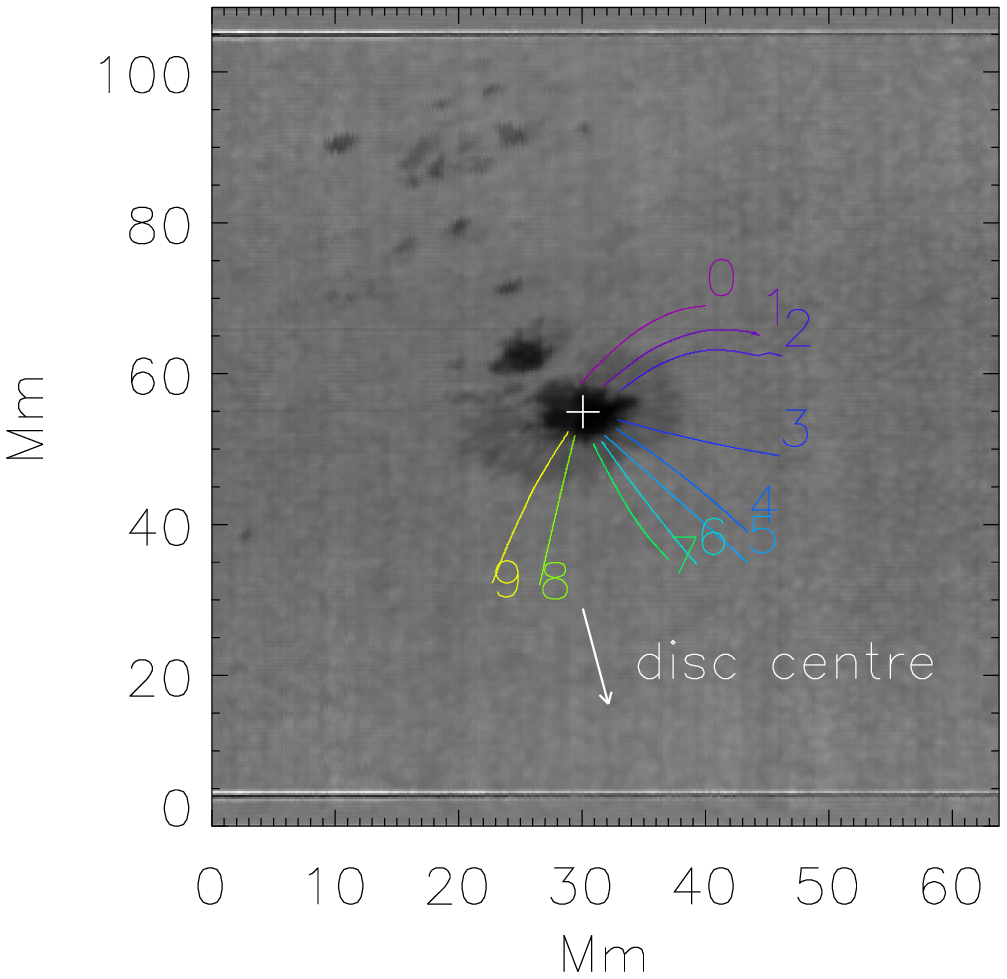}}
\resizebox{1.5cm}{!}{\includegraphics{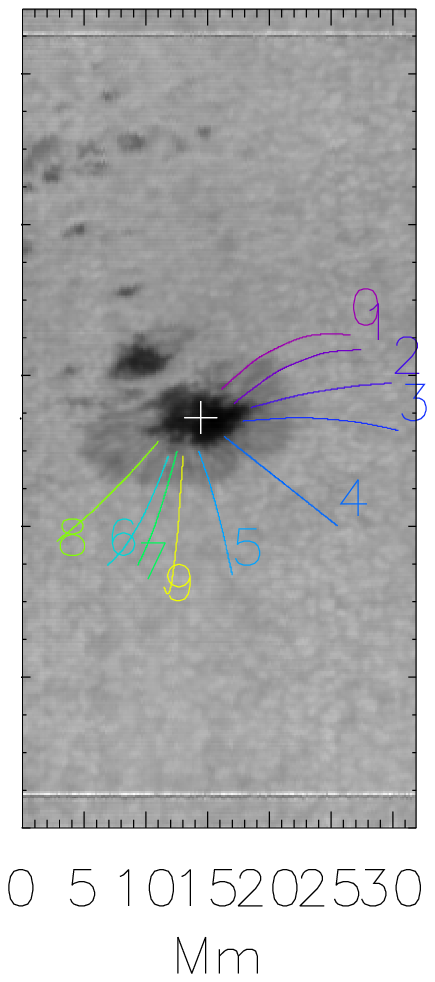}}
\resizebox{1.5cm}{!}{\includegraphics{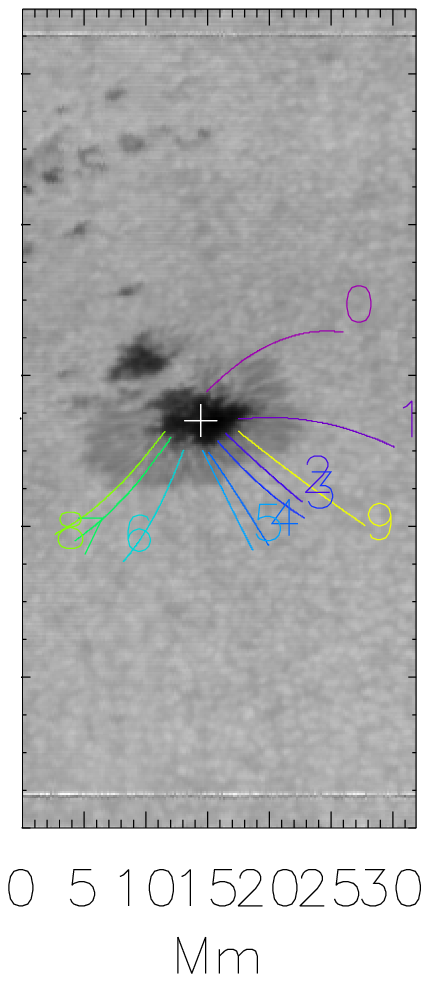}}
\resizebox{1.5cm}{!}{\includegraphics{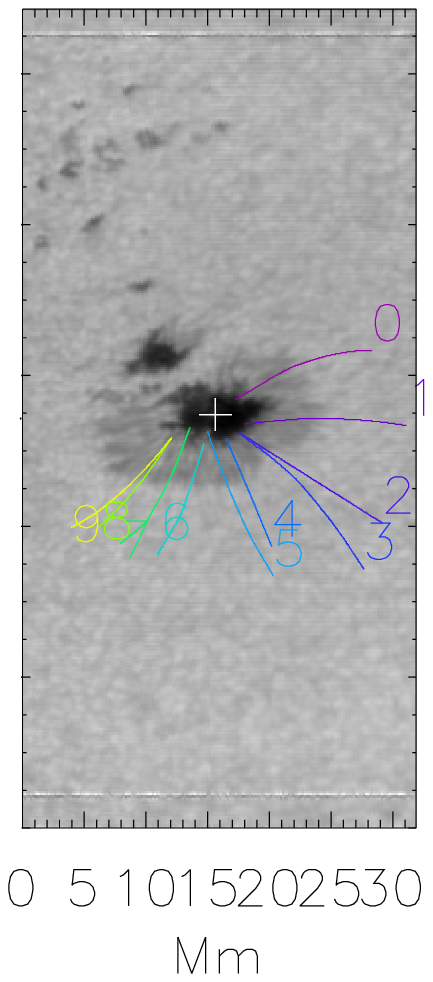}}
\resizebox{1.5cm}{!}{\includegraphics{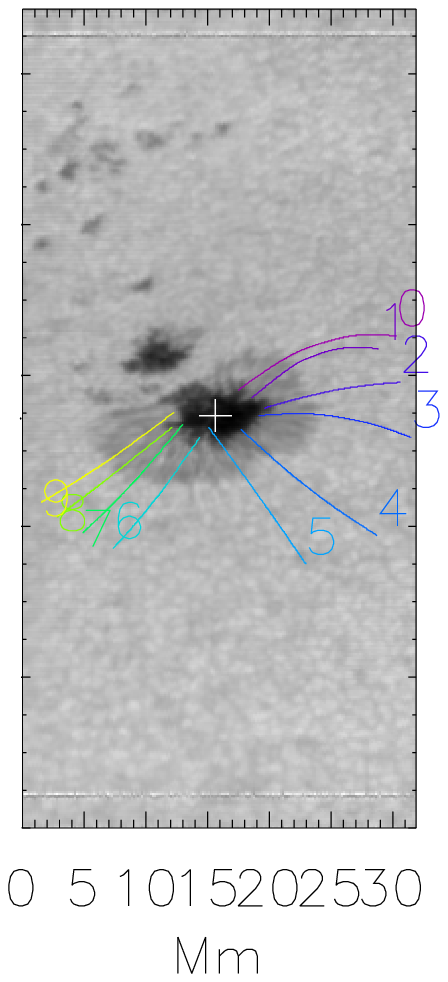}}
\resizebox{1.5cm}{!}{\includegraphics{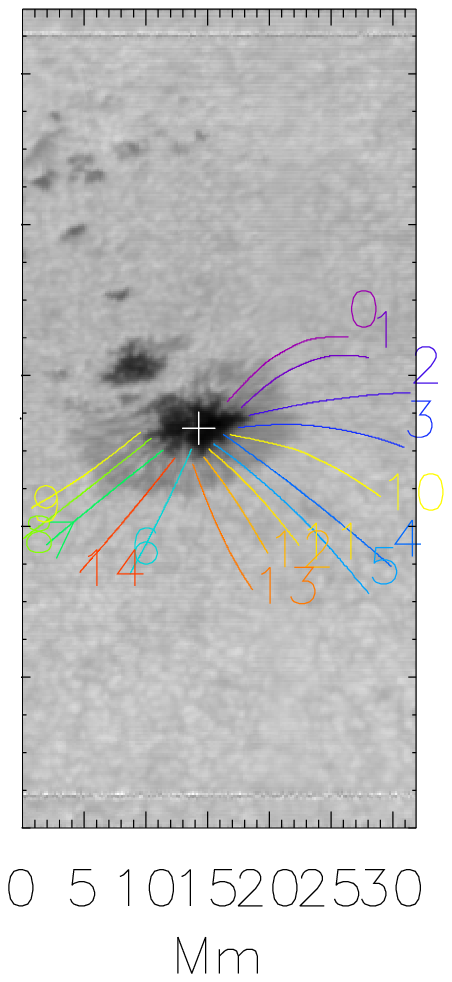}}
\resizebox{1.5cm}{!}{\includegraphics{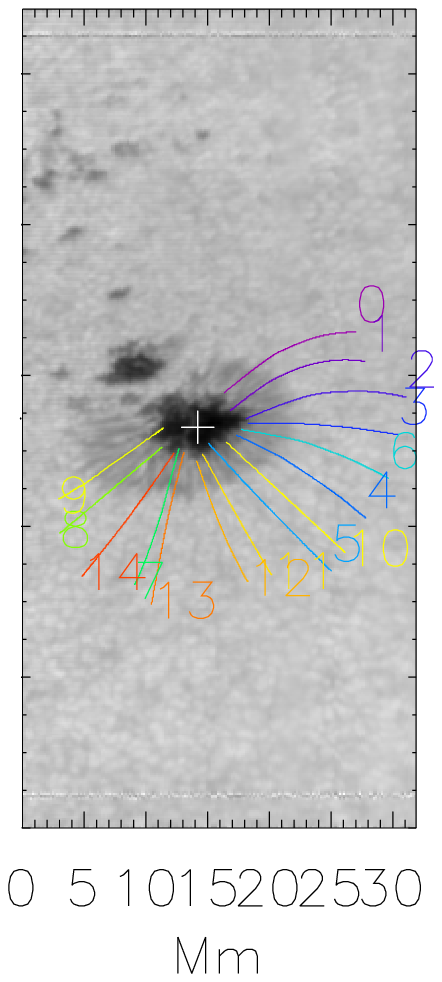}}
\resizebox{1.5cm}{!}{\includegraphics{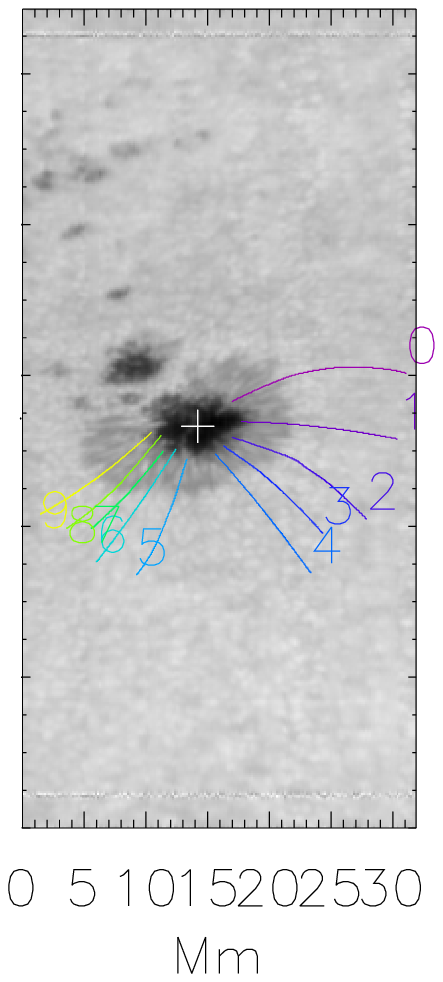}}
\resizebox{1.5cm}{!}{\includegraphics{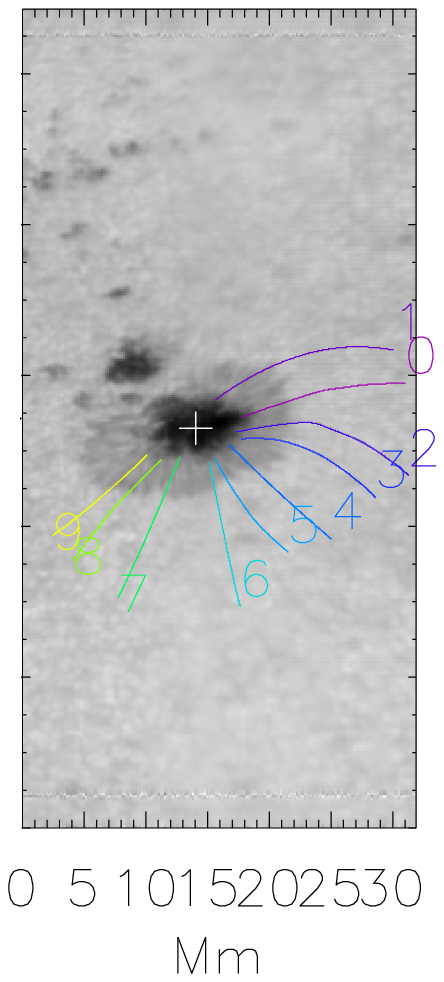}}$ $\\$ $\\$ $\\
\caption{Overview of the nine maps taken on 03/08/2013 from UT 15:24-18:13. Bottom to top: continuum intensity, photospheric velocity, line-core intensity and line-core velocity of \ion{Ca}{ii} IR at 854.2\,nm and mask of fast downflows. Colored lines indicate the locations of fibrils labeled from 0-9 (0-14) in each scan. The white pluses in the bottom row indicate the center of the sunspot. The different colored symbols in the second row from the bottom indicate the locations of maximal velocities (diamonds) and maximal line-core intensity (pluses) along each fibril. The maps have not been resized to have square pixels. (Please magnify for better visibility of small details or labels)}\label{fig2}
\end{figure*}

The photospheric spectra at 1565\,nm were inverted with the \textit{Stokes Inversion based on Response Functions} code \citep[SIR;][]{cobo+toroiniesta1992} including eight lines with known transitions inside the observed wavelength range. These results on the photospheric magnetic field will be used in a subsequent study. The intensity profiles of \ion{Ca}{ii} IR were inverted under the assumption of local thermodynamic equilibrium (LTE) with the \textit{CAlcium Inversion using a Spectral ARchive} code \citep[CAISAR;][]{beck+etal2013,beck+etal2015}. We used the non-LTE correction curve for the penumbra from \citet{beck+etal2015} for the scaling of the LTE temperatures where indicated.

We manually identified chromospheric fibrils in the nine maps using the line-core intensity and velocity of \ion{Ca}{ii} IR. Each fibril was marked by a set of six spatial positions with a subsequent fit of a second-order polynomial to these points. The first point for each fibril track was chosen to be at about the border between umbra and penumbra along the continuation of the fibril into the sunspot. The fibrils tracks were selected to capture the most prominent, fastest and isolated downflow patches near the sunspot in each map and to trace the corresponding flow fibril outwards from the spot as far as possible. The outermost point of the curves was set to continue the fibril track for about 5$^{\prime\prime}$ beyond the last clear signature of the fibril itself. We selected 15 fibrils in the 6th and 7th map and ten fibrils in all others (100 in total; see Figure \ref{fig2}). 

The locations of downflow patches were determined by a threshold in the chromospheric line-of-sight (LOS) velocity. All neighbouring spatial positions with a LOS velocity above the threshold were attributed to an individual downflow patch (top row of Figure \ref{fig2}). Most of the manually identified fibrils have an associated patch in these masks, while some patches correspond to fibrils that we did not select. The total number of downflow patches was comparable to that of individual fibrils.

\begin{figure*}
\centerline{\resizebox{17.2cm}{!}{\hspace*{.5cm}\resizebox{4.5cm}{!}{\includegraphics{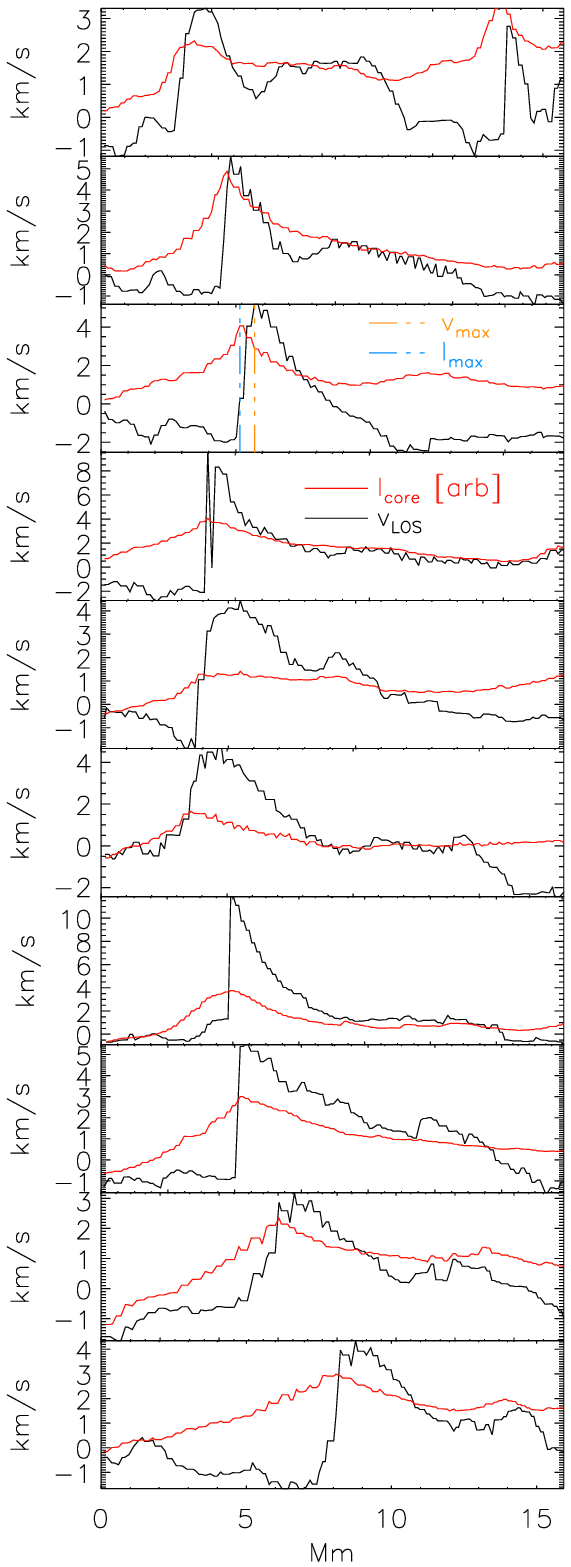}}\hspace*{1.5cm}\resizebox{10cm}{!}{\includegraphics{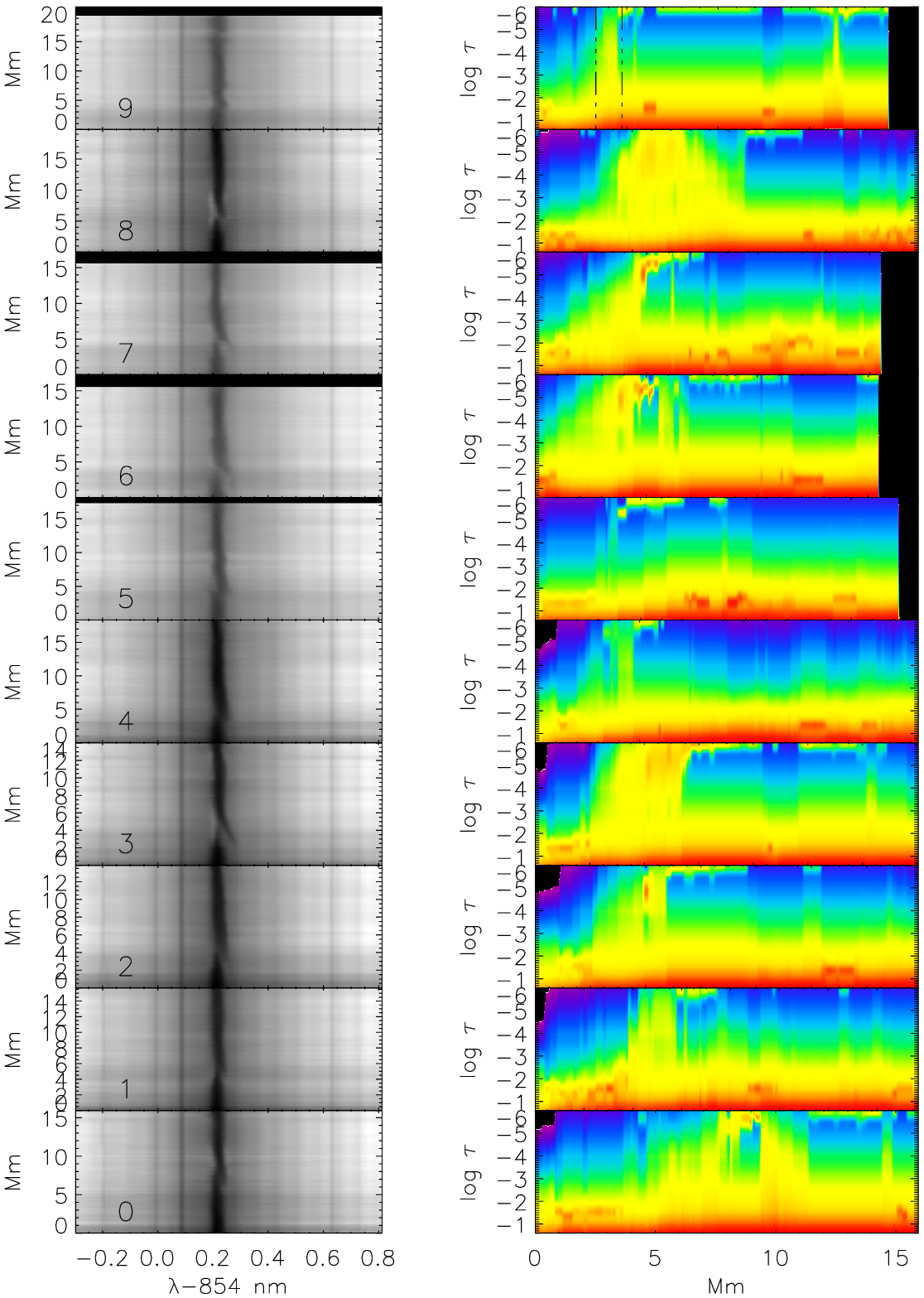}}\hspace*{.5cm}\resizebox{1.5cm}{!}{\includegraphics{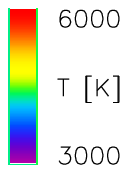}}}}$ $\\
\caption{Atmospheric properties and spectra along the ten fibrils in the second map. The numbers given in the middle column correspond to those in the second column of the bottom row of Figure \ref{fig2}. Left column: line-core intensity (red lines) and chromospheric velocity (black lines) along the fibrils. Middle column: intensity spectra. Right column: temperature stratifications. Uniform black areas in the spectra or the temperature were beyond the end of the cut along the fibril. The umbra of the sunspot is at the left in the left/right column and at the bottom in each panel in the middle column. The two dash-dotted vertical lines in the third panel from the top in the left column indicate the locations of maximum velocity and intensity for this fibril. The dash-dotted vertical lines in the uppermost panel of the right column mark the determination of the width of the fibril.}\label{fig3}
\end{figure*}

Given the fibril locations, we extracted the spectra, $I_c$, $I_{\rm core}$, $v_{\rm core}$, $v_{\rm phot}$ and the temperature stratifications T(log $\tau$) from the LTE inversion along each fibril. We determined the location of maximal velocity and maximal line-core intensity along each fibril, but enforced that the positions corresponded to the location closer to the sunspot in case that there were multiple local maxima along the fibril. For the downflow patches, we determined the average velocities and intensities and the maximal temperature at each optical depth. The area of each downflow patch was derived from all pixels pertaining to it. 
\section{Results} \label{secres}
Figure \ref{fig2} shows the SPINOR observations in photospheric and chromospheric quantities. The selected fibril tracks are indicated by colored lines. The sunspot only had a regular penumbra and chromospheric fibrils on the side opposite to both the new flux emergence and the following polarity. A large fraction of the locations of maximal velocities and line-core intensity along each fibril that are indicated by colored symbols in the second row from the bottom of Figure \ref{fig2} is inside the penumbra, while the rest is at least close to the sunspot.

\begin{figure*}
\centerline{\resizebox{17.2cm}{!}{\hspace*{.5cm}\resizebox{4.5cm}{!}{\includegraphics{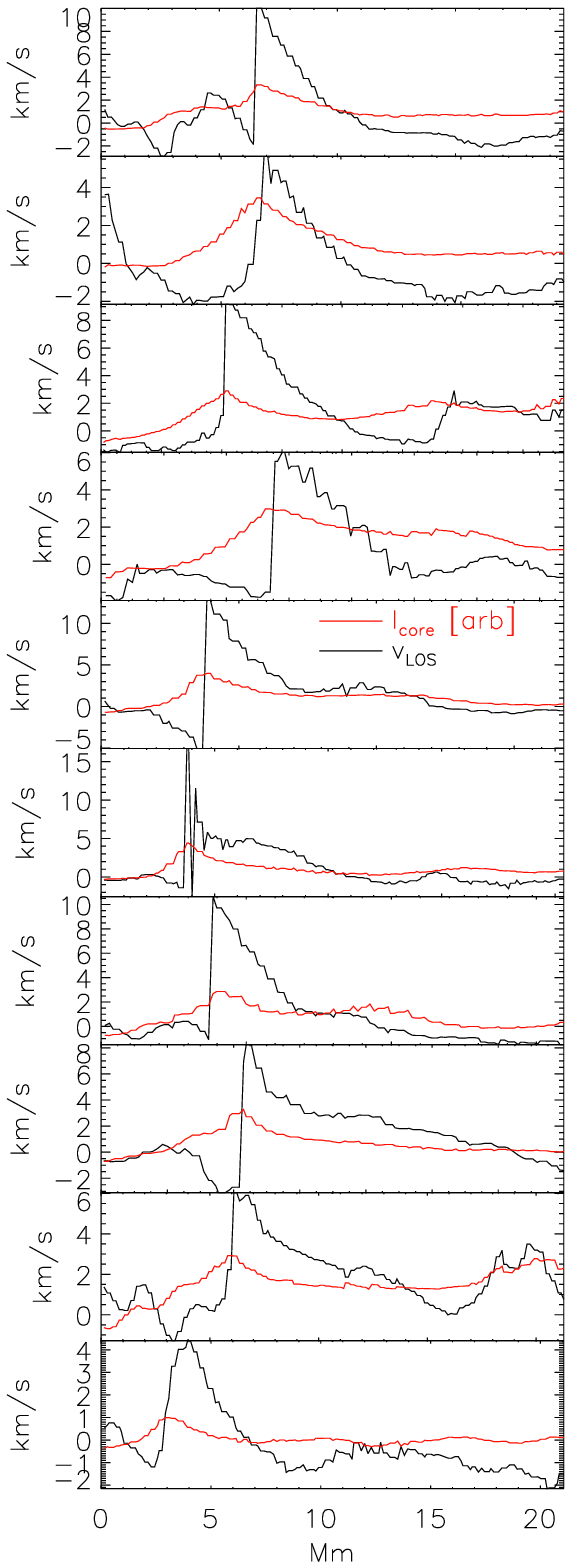}}\hspace*{1.5cm}\resizebox{10cm}{!}{\includegraphics{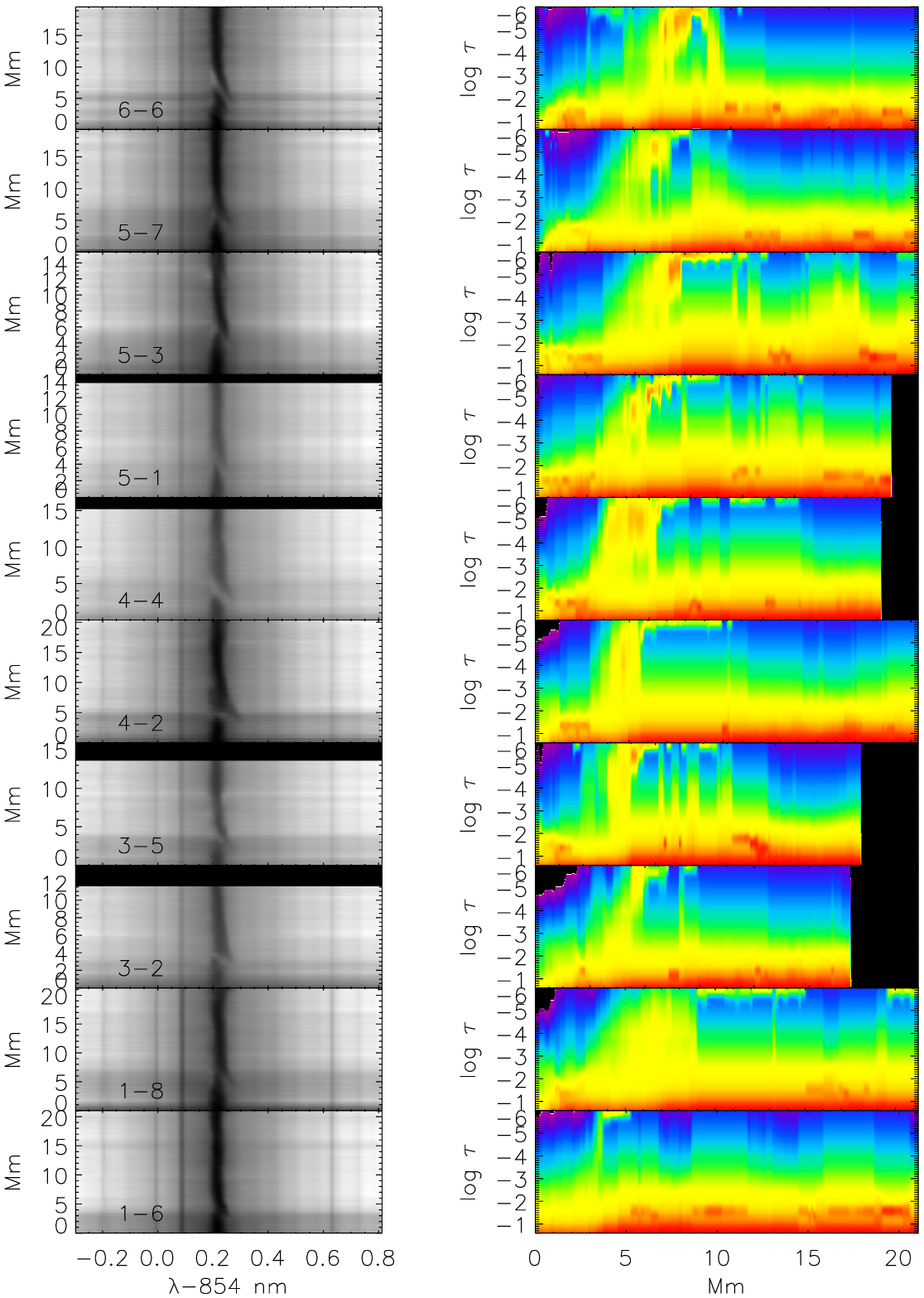}}}}$ $\\
\caption{Same as Figure \ref{fig3} for a selection of fibrils from different maps. The numbers A-B in the middle column denote the map (A) and the fibril in the map (B) corresponding to the labels in the bottom row of Figure \ref{fig2}.}\label{fig4}
\end{figure*}
\subsection{Thermodynamic parameters along individual fibrils}
Figure \ref{fig3} shows the thermodynamic properties along all ten fibrils in the second map, while Figure \ref{fig4} shows the same for a selection of fibrils from different maps. The line-core intensity (red line in the left columns of Figures \ref{fig3} and \ref{fig4}) smoothly increases towards the location of the fastest downflow for each fibril. It peaks slightly closer to the sunspot that is located to the left in the plots than the velocity. The peak in the line-core intensity is roughly symmetric around its maximum. The line-core velocity increases fast when approaching the sunspot from the outside and then drops abruptly to about zero over a short distance of less than 1\,Mm in most cases. The maximal chromospheric LOS velocities are in the range of 4--15\,km\,s$^{-1}$. 

The pattern of intensity and velocity is also clearly visible in the spectra along the fibrils (middle columns of Figures \ref{fig3} and \ref{fig4}). The line core of \ion{Ca}{ii} IR shows clear and large Doppler shifts with an increase in the intensity starting already further away from the sunspot. The intensity in the line wing usually also increases near the downflows over the full spectral range shown. The line core at the locations beyond the maximal flow speed reverts again to about zero Doppler shift.  

The temperature stratifications along the fibrils (right columns of Figures \ref{fig3} and \ref{fig4}) show temperature enhancements over most of the optical depth range at the downflow locations. The temperature enhancement over the surroundings disappears in the upper photosphere at about log $\tau \approx -2$ to $-3$. For the majority of the fibrils, we find a tube-like structure of 1-3 Mm width close to the fastest downflow that makes an angle of 0--30 degrees to the LOS that corresponds to the y-axis in the temperature images. Most fibrils can only be consistently traced for 2--5\,Mm from the downflow point in the temperature plots. The majority rises beyond log $\tau = -6$ out of the formation height of the \ion{Ca}{ii} IR line. A few fibrils show an intermittent temperature enhancement close to log $\tau = -6$ far away from the downflow point that could indicate that there is only a weak signature of the fibril left at these distances. From the corresponding temperature plots for all of the fibrils in all maps, only about 10\,\% show a structure that returns to the photosphere again as in the bottom (top) right panel of Figure \ref{fig3} (Figure \ref{fig4}). 

\begin{figure*}
\centerline{\resizebox{14cm}{!}{\includegraphics{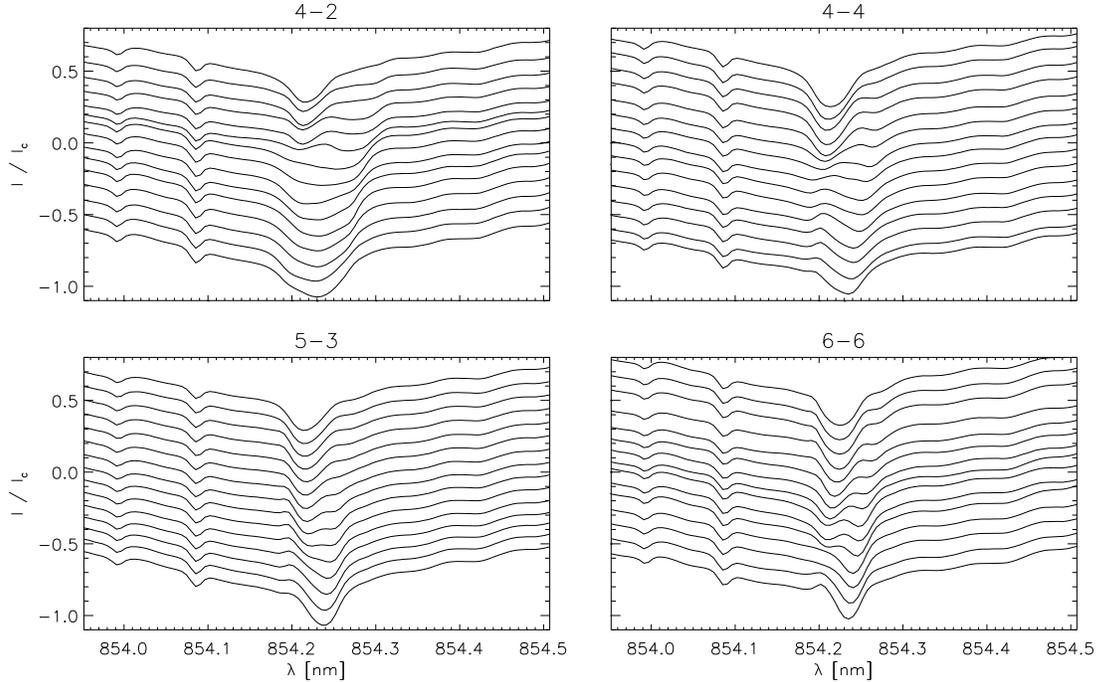}}}$ $\\
\caption{Example spectra around the location of largest Doppler shifts for individual fibrils. The numbers A-B on top of each panel denote the map (A) and the fibril in the map (B) corresponding to the labels in the bottom row of Figure \ref{fig2}. The sunspot is located at the top in each panel.}\label{fig5}
\end{figure*}

The maximal Doppler shifts near the downflow points are rather large so that the spectra (middle columns of Figures \ref{fig3} and \ref{fig4}) give the impression of having two separate components in some cases. We therefore checked whether our automatic procedure for the derivation of the line-core velocity actually works correctly for such profiles because it determines the location with maximal line depth inside the profile. Figure \ref{fig5} shows individual spectra of four fibrils around the downflow points. The red-shifted flow component is usually the dominant component in the spectra over some range away from the sunspot, but then converts into being a weak line satellite of an un-shifted, stronger absorption component. Wherever the un-shifted component is deeper, the automatic determination yields its Doppler shift instead of that of the flow component. 

We therefore manually traced the weak Doppler-shifted component in a subset of fibrils to determine the maximal Doppler shift up to the location where the weak component disappears completely. The velocity values from the automatic derivation for the same profiles usually underestimated the true velocity by a factor of about two. Almost all of the selected downflow patches in the velocity maps and also the flow speeds along the fibril thus actually are close to or slightly super-sonic with typical speeds above 8\,km\,s$^{-1}$. The manual determination also shifted the location of the maximal velocity more towards the sunspot, which gives a better match to the location of maximal line-core intensity. The latter quantity also suffers from the fact that the increasing Doppler-shift of the second component raises the line-core intensity -- defined as the minimal intensity in the profile -- because the absorption profile of the second component moves to the red.

\subsection{Average fibril properties}
We then averaged the thermodynamic properties over all of the 100 fibrils. We used the location of maximal velocity along each fibril as the common reference point to align the different fibril tracks with each other. 

\begin{figure}
\resizebox{8.8cm}{!}{\includegraphics{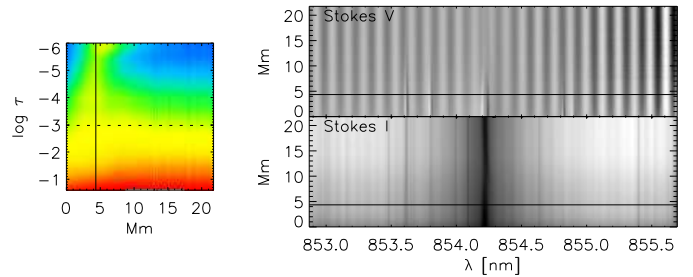}}\\$ $\\
\caption{Temperature (left panel) and spectra (right panel) averaged over all fibrils. The vertical black line in the left panel indicates the location of maximal velocity while the horizontal dashed line marks log $\tau =-3$. The horizontal black lines in the Stokes I (bottom right panel) and Stokes V spectra (top right panel) indicate the location of maximal velocity.}\label{fig6}
\end{figure}

Figure \ref{fig6} shows the temperature stratifications and the Stokes I and V spectra along the average fibril. The structure has been somewhat smoothed out through the averaging, but the description given for individual fibrils above is still valid. At the location of the fastest downflow at about 3--5\,Mm, the average temperature shows a barely inclined, thin (1--2\,Mm) tube with a local temperature increase. At an optical depth of about log $\tau \approx -3$, this temperature enhancement is no longer apparent. The temperature enhancement rises up to log $\tau = -6$ with increasing distance from the sunspot, but can only be tracked there for about 5\,Mm from the downflow point. 

The average Stokes I spectra do not show the characteristic intensity and Doppler shift pattern very prominently because the full spectral range of the profiles is displayed to cover also the photospheric line blends. The Stokes V spectra are more revealing in this case. The average location of the maximal flow velocity (black horizontal line in the right panel of Figure \ref{fig6}) intersects clear circular polarization signal in both photospheric and chromospheric lines, but is still not in the region of the strongest polarization signal. 

\begin{figure*}
\centering
\resizebox{17cm}{!}{\includegraphics{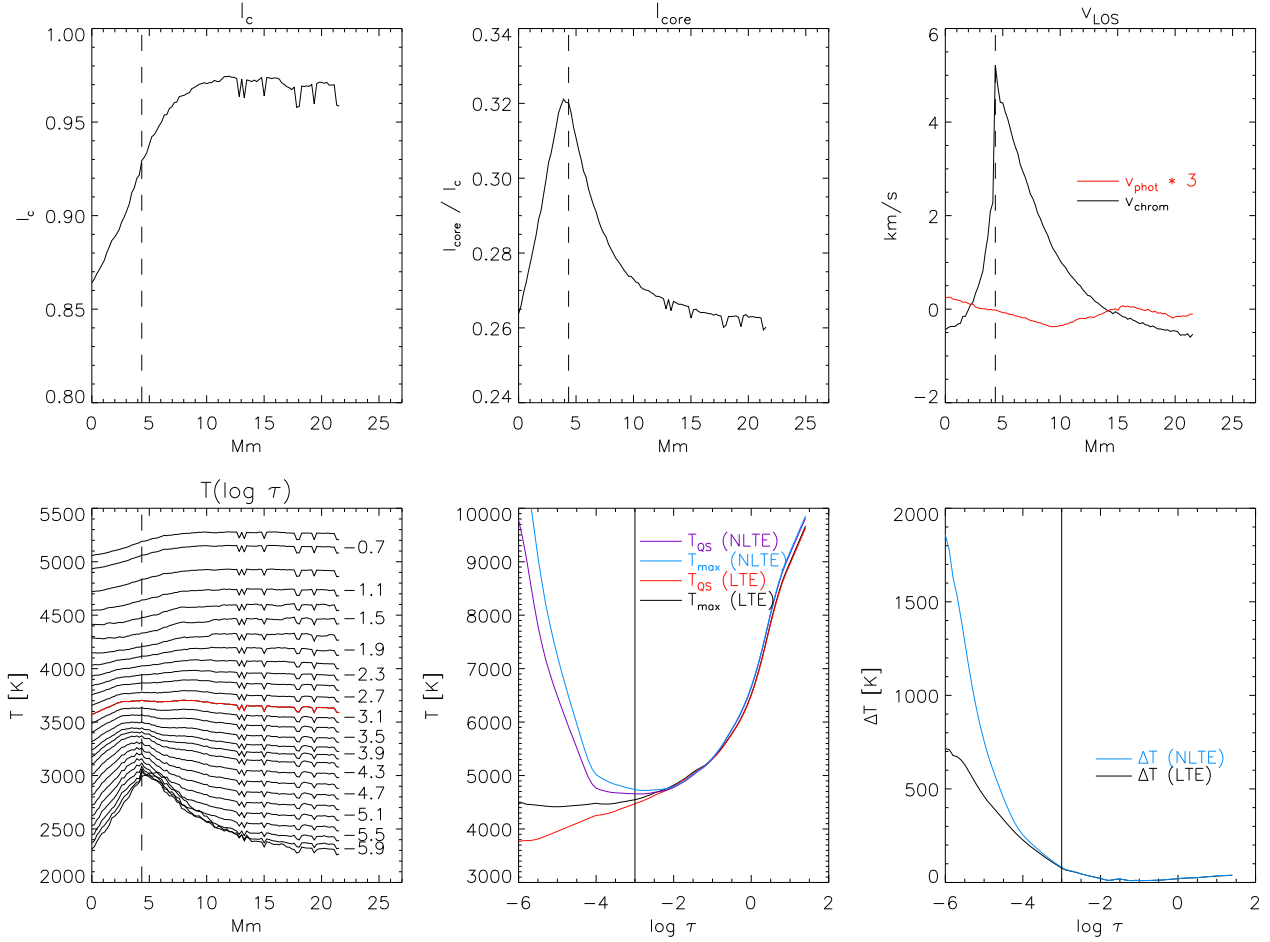}}
\caption{Thermodynamic properties of the average fibril. Top row, left to right: continuum intensity, line-core intensity and line-core velocity of \ion{Ca}{ii} IR at 854.2\,nm along the average fibril. The vertical dashed line indicates the location of maximal velocity. The red line in the upper right panel shows the photospheric velocity multiplied by a factor of 3 for better visibility. Bottom row, left: temperature along the averaged fibril at different optical depths given at the right end of each curve. The red line for T(log $\tau = -3$) is flat at the downflow point. Bottom row, middle: temperature as a function of optical depth. Black/blue lines: maximal temperature in the average fibril in LTE/NLTE. Red/purple lines: temperature in a QS reference area in LTE/NLTE. The vertical black line marks log $\tau = -3$. Bottom row, right: difference between maximal and QS temperature in LTE/NLTE (black/blue lines).}\label{fig7}
\end{figure*}

The top row of Figure \ref{fig7} shows the continuum intensity, the line-core intensity and the line-core velocity after averaging over the fibrils. The continuum intensity (top left panel of Figure \ref{fig7}) confirms that the fastest downflows are on average located inside the penumbra where $I_c$ drops below unity. The peak in the line-core intensity is roughly symmetric, while the line-core velocity (top right panel of Figure \ref{fig7}) shows a fast increase towards the sunspot with a sharp drop from the peak velocity of 5\,km\,s$^{-1}$ to about 2\,km\,s$^{-1}$ on less than 1\,Mm. There is a less steep decrease from 2\,km\,s$^{-1}$ to about zero over the next 3\,Mm.

The bottom row of Figure \ref{fig7} shows the properties of the average temperature stratifications. The temperature (lower left panel) is enhanced at the downflow location from log $\tau = -6$ to about log $\tau = -3$. At lower atmospheric layers, the temperature is actually lower than further away from the sunspot, reflecting the location in the penumbra. To quantify the temperature enhancement, we averaged the inversion results over a quiet Sun (QS) reference area in the FOV. The middle panel of Figure \ref{fig7} compares the temperature in the QS (red line) with the maximal value at each layer of optical depth (black line), while the right panel shows their difference. The maximal value of T$_{\rm max}$(log $\tau$) along the average fibril is not always exactly at the location of the fastest flow (vertical dotted line in the lower left panel of Figure \ref{fig7}), but the difference is usually very small. In the LTE results, the downflows are hotter than the QS stratification by 200\,K at log $\tau = -4$ and by 700\,K at log $\tau = -6$ (bottom right panel of Figure \ref{fig7}). After application of the non-LTE correction curve for the penumbra, the difference increases to about 1800\,K at log $\tau = -6$, while the value at log $\tau = -4$ does not change much.

\begin{figure*}
\begin{minipage}{12cm}
\resizebox{12cm}{!}{\includegraphics{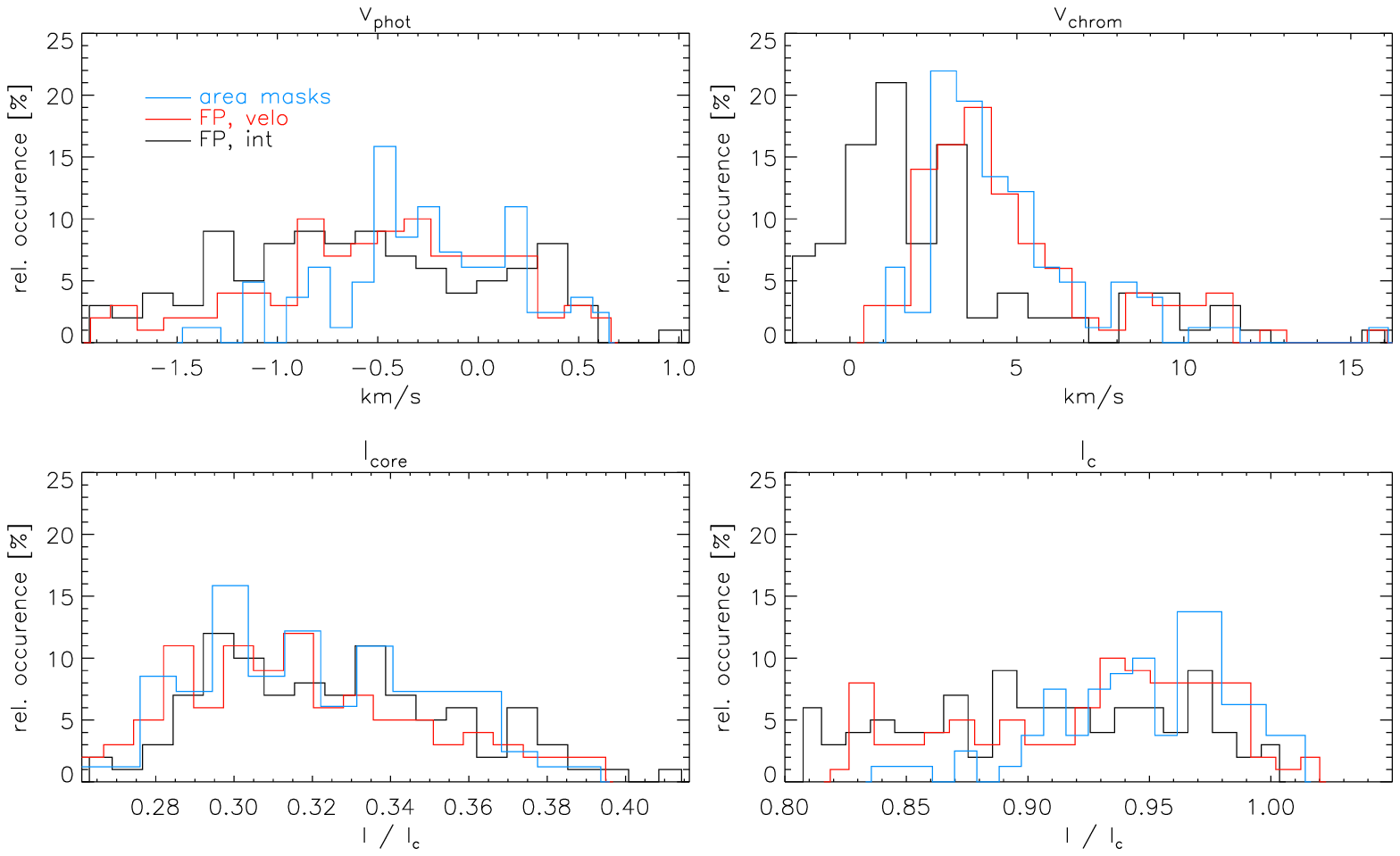}}
\end{minipage}\begin{minipage}{6cm}
\resizebox{6.25cm}{!}{\includegraphics{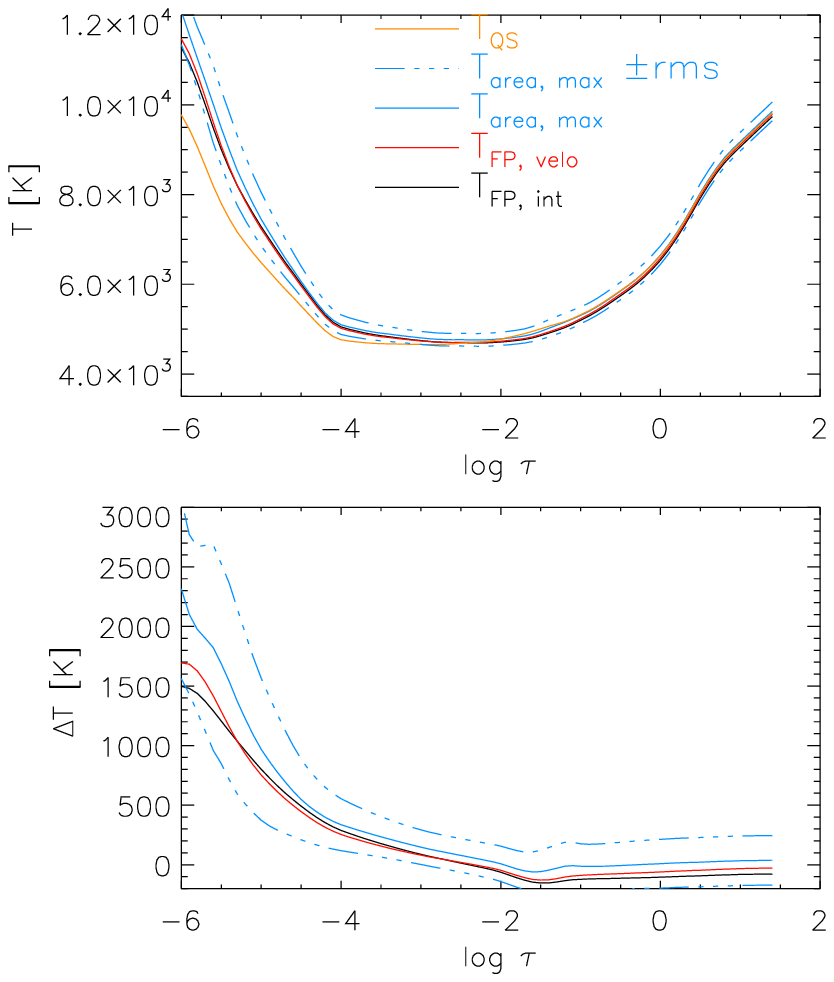}}\vspace*{.5cm}
\end{minipage}
\caption{Statistics of thermodynamic properties at the downflow locations. Left two columns: histograms of the line-core intensity (bottom left), continuum intensity (bottom middle), photospheric velocity (top left) and chromospheric velocity (top middle). Black/red lines: values at the locations of maximal intensity/velocity. Blue lines: values averaged over the area of the downflow patches in the masks (top row of Figure \ref{fig2}). Right top: average temperature stratifications at the locations of maximal intensity/velocity after application of the NLTE correction (black/red lines), in the QS reference area (orange line), average maximal temperature in each downflow patch (solid blue) and the 1-$\sigma$ variance of the latter (dash-dotted blue line). Bottom right: same as above for the difference to the temperature in the QS reference area.}\label{fig9}
\end{figure*}
\subsection{Statistics of thermodynamic parameters of fibrils and downflow patches}
Figure \ref{fig9} shows a comparison of the statistics of values taken at the locations of maximal velocity or intensity along the fibrils and the corresponding values inside the downflow patches that were defined through a threshold in the velocity maps. The histograms of chromospheric and photospheric intensities and velocities from the different areas are rather similar. The photospheric velocity shows a small redshift of about 0.4\,km\,s$^{-1}$, while the chromospheric velocity covers a range from 0 to 15\,km\,s$^{-1}$ without any correction for the underestimation by the automatic method. The histograms of the line-core intensity are nearly identical, while the continuum intensity is predominantly smaller than unity. 

The temperature stratifications at the different locations show a similar enhancement over the QS as the stratification for the average fibrils in Figure \ref{fig7}. For the downflow patches, we selected the maximal temperature inside each patch at each layer of optical depth and then averaged these values over all patches. The temperature increase over QS conditions exceeds 2000\,K at log $\tau = -6$. The corresponding standard deviation across all patches is about $\pm$ 200\,K from log $\tau = 0$ to log $\tau = -4$, while it increases to about $\pm$ 700\,K at log $\tau = -6$.

\subsection{Statistics of spatial properties}
\begin{figure}
\resizebox{6cm}{!}{\includegraphics{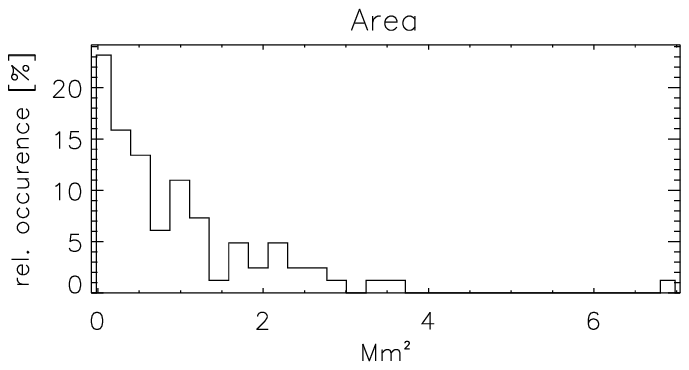}}\\
\resizebox{6cm}{!}{\includegraphics{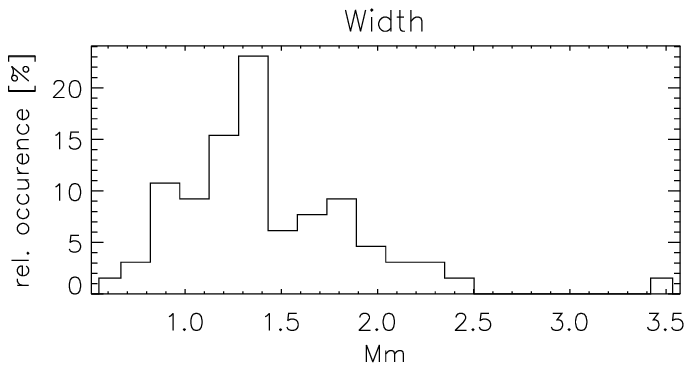}}\\
\resizebox{6cm}{!}{\includegraphics{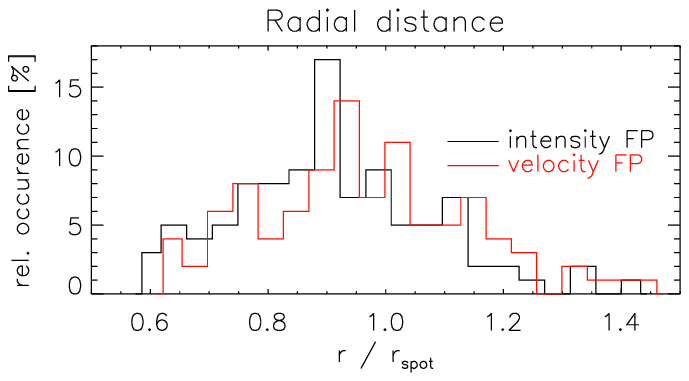}}
\caption{Statistics of spatial properties. Bottom panel: radial distance of locations of maximal intensity/velocity (black/red line) from the center of the sunspot in fractional spot radii. Middle panel: histogram of the width of the fibrils in the 2D temperature plots along the fibrils (cf.~upper rightmost panel of Figure \ref{fig3}). Top panel: histogram of the area of the downflow patches in the masks.}\label{fig10}
\end{figure}

Figure \ref{fig10} shows the histograms of the area of downflow patches, the width of the fibrils and the radial distance of the locations of maximal velocity and intensity from the center of the sunspot. The downflow patches (cf.~the top row of Figure \ref{fig2}) are usually thin and somewhat elongated in the radial direction. Typical areas are below 2 Mm$^2$ (top panel of Figure \ref{fig10}). 

The width was determined manually in the temperature plots along individual fibrils (cf.~top right panel of Figure \ref{fig3}), but only for a subset of the fibrils. In some cases, especially for the broader and more diffuse structures, it was not really possible to define an outer border of the flow fibril. There might thus be some slight bias towards lower widths. With this caveat, the typical value for the width of the flow fibrils at the downflow points is 1--2.5\,Mm. Some of the variation in the width will also come from the fact that the LOS intersected different fibrils around the sunspot at an different azimuth angle. 

Finally, the histogram of the radial distance in the bottom panel of Figure \ref{fig10} confirms that the majority of the downflow points are located inside of or close to the outer boundary of the penumbra. The radial distance in fractions of the sunspot radius is from 0.6--1.4. We note that the smallest radius values are still outside of the umbra. There is no obvious case of similar fast chromospheric downflows ending inside the umbra while several are clearly located outside of the sunspot (cf.~Figure \ref{fig2}). 

\section{Summary }\label{secsumm}
We observed a sunspot in active region NOAA 11809 at a heliocentric angle of 30$^\circ$ with SPINOR at the DST. We acquired nine maps of spectra of the \ion{Ca}{ii} IR line at 854.2\,nm under good seeing conditions. We manually identified and traced 10--15 chromospheric fibrils per map in or near the undisturbed segment of the sunspot that showed a penumbra. Those chromospheric velocity and intensity fibrils originated from the mid penumbra up to locations outside the sunspot and extended further outward into the super-penumbral canopy while showing the inverse Evershed effect. 

The patches of maximum flow velocities were primarily found near the outer penumbral boundary from 0.6 to 1.4 spot radii, with a peak of the distribution inside the sunspot at 0.9 radii. The chromospheric velocities of 3--15\,km\,s$^{-1}$ were about one order of magnitude higher than the photospheric velocities. The thermal inversion results showed that the flow channels are not horizontal near their downflow points, but are instead inclined to the local vertical by 30--60 degrees. The end points of the flow channels were found to show an abrupt drop in velocity and an enhancement of intensity that is characteristic for the presence of a shock front. On average, the shock front was located at about 5\,Mm from the outer umbral boundary and occured at heights from log\,$\tau \sim -2.7$ to $-6$. The temperature in the shock was increased relative to the quiet Sun by about 200 and 2000\,K at these two optical depths. The shock locations cannot be identified in the maps of the continuum intensity and the temperature of layers below log\,$\tau \sim -2.7$. The lateral width of the fibrils in spatial maps and their extent in the vertical temperature stratifications was from below 1 up to 2.5\,Mm, while the area of the downflow patches was from almost zero to 4\,Mm$^2$.

\begin{figure}
\resizebox{8.cm}{!}{\includegraphics{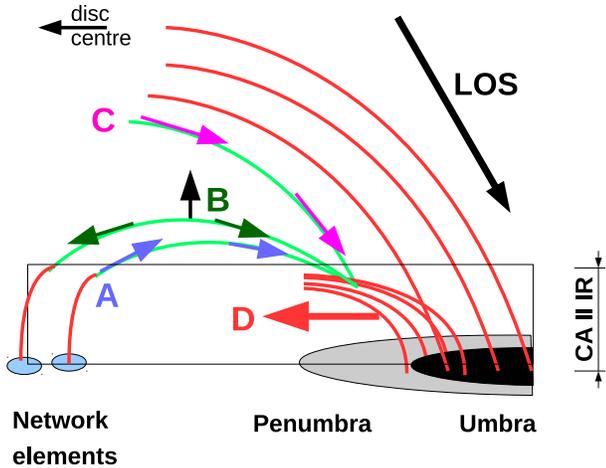}}
\caption{Sketch of different flow scenarios. A: Inverse Evershed flow (IEF) through a siphon flow mechanism. B: IEF caused by mass shedding of rising field lines. C: IEF through coronal rain. D: photospheric Evershed flow. Magnetic field lines in cyan are those that are most relevant for the IEF. The black rectangle indicates the formation height of the \ion{Ca}{ii} IR line at 854.2\,nm from the continuum-forming layers to a height of about 1\,Mm.} \label{fig_sketch}
\end{figure}

\section{Discussion}\label{secdisc}
The mass motions along the fibrils, which are assumed to trace the magnetic field lines in the super-penumbra of the sunspot, could result from a variety of different phenomena that are illustrated in Figure \ref{fig_sketch}. Configuration "A" represents a siphon flow \citep[e.g.,][]{meyer+schmidt1968,cargill+priest1980,thomas1988,degenhardt1989,uitenbroek+etal2006} that originates in a magnetized region in the super-penumbra, either a moving magnetic feature in the sunspot moat or the network at its boundary, and ends near or in the penumbra. The flow in this case is maintained by the pressure difference at the two photospheric end points that are connected by the fibrils. It is characterized by an upflow at the end point of lower field strength and a downflow at the one with higher field strength \citep[e.g.,][]{bethge+etal2012}. In case “B”, a rising flux bundle drains plasma along its length. That causes downflows at both ends and upward velocities at the middle of the fibril \citep[e.g.,][]{gomory+etal2010}. Configuration "C" represents the scenario of coronal rain \citep[see, e.g.,][and references therein]{antolin+etal2012}, in which cool, dense coronal condensations slide down along guiding magnetic field lines. Such flows commonly show very high speeds of up to 80 km\,s$^{-1}$ and are often intermittent in time \citep{visservandervoort2012,ahn+etal2014}. Our flow pattern with rather steady, uniform downflows of about sound speed over a connected segment of the sunspot matches best to the siphon flow scenario, as upflows or multiple times super-sonic speeds are absent.

Earlier measurements of the inverse Evershed flow using a coarse wavelength sampling in the H$\alpha$ line showed large spatio-temporal variations in the flow speed \citep{georgakilas+etal2003}. This could result from the fact that the flow often only shows up as weak line satellite (Figure \ref{fig5}) that would be impossible to isolate at low spectral resolution. The flow speeds reported here are consistent with the flow pattern observed in H$\alpha$ and \ion{C}{iv} with a sharp decrease in the downflow velocity a few arcseconds outside the penumbra \citep{alissandrakis+etal1988,georgakilas+etal2003}. The location of maximal downflow should vary with the heliocentric angle of the observations. A study of the center-to-limb variation of the inverse Evershed flow should be able to reveal this effect. The velocity profiles along the fibrils or their average in our observations with a decelaration to nearly zero over less than 1\,Mm matches the shape of the flow speed shown in \citet[][their Figure 3]{bethge+etal2012} for a siphon flow in \ion{He}{i} at 1083\,nm that shows significantly larger speeds up to 40 km\,s$^{-1}$. The average flow profile also matches well the values predicted by \citet[][middle panel of their Figure 3]{montesinos+thomas1993aa} for a photospheric siphon flow of critical speed.

The formation height of the \ion{Ca}{II} line at 854.2 nm samples the solar atmosphere from continuum-forming layers to where the chromospheric plasma turns from high to low plasma-$\beta$ \citep{pietarila2007}. The different siphon flow solutions in \citep{thomas1988} or \citet{montesinos+thomas1993aa} that attain super-sonic speeds predict the occurence of a shock front when the super-sonic chromospheric or upper-photospheric flow encounters the denser lower atmosphere. Such shock fronts are clearly indicated by the brightenings in the line-core intensity of the \ion{Ca}{II} line at 854.2 nm, while the inversion of the spectra allowed us to also localize their extent through the atmosphere (log $\tau \sim -3$ to $-6$) and the corresponding increase in temperature (200--2000\,K). The lower boundary of the shocks corresponds to upper photospheric layers, while the temperature increase is comparable to that in shocks from acoustic waves \citep{beck+etal2013a}. 

In total, the flow and thermal patterns of individual chromospheric flow and intensity fibrils observed here resemble most a siphon flow in a magnetic flux tube in a stratified atmosphere, where the flow velocity increases towards the downstream end to a point where it turns super-sonic and decelerates abruptly because of forming a shock front. The spatial location of the shock is in or just outside the outer penumbral boundary, while in the current data the outer end of the fibril cannot be identified.

The inversion with CAISAR \citep{beck+etal2015} also allowed us to determine the angle of the inverse Evershed flow near the downflow points -- under the assumption that the temperature enhancements trace the flow fibrils -- with spatial resolution and without recourse to any additional assumptions on the flow geometry. In a follow-up paper, we plan to study the flow angle after conversion to the local solar surface in relation to the properties of the magnetic field at photospheric and chromospheric heights. The latter can be derived from the analysis of the spectropolarimetric observations at 1565\,nm and in \ion{Ca}{ii} IR or magnetic field extrapolations.

\section{Conclusions}\label{secconcl}
We have studied the thermodynamic structure and properties of the solar atmosphere along chromospheric intensity and velocity fibrils in the super-penumbra of a sunspot that harbor the inverse Evershed effect. We find that the properties match best to a siphon flow that attains super-sonic speed and ends in a shock front inside the penumbra of the sunspot. The flows are inclined by 30--60 degree to the local vertical near the downflow points. 

\begin{acknowledgements}
This work is supported by NSF grant 1413686. The Dunn Solar Telescope at Sacramento Peak/NM was operated by the National Solar Observatory (NSO). NSO is operated by the Association of Universities for Research in Astronomy (AURA), Inc.~under cooperative agreement with the National Science Foundation (NSF). HMI data are courtesy of NASA/SDO and the HMI science team. 
\end{acknowledgements}
\bibliographystyle{aa}
\bibliography{references_luis_mod}

\end{document}